\documentclass{ws-ijqi}

\newcommand{\floor}[1]{ {\lfloor #1 \rfloor } }
\newcommand{\ket}[1]{ {\left| #1 \right\rangle} }
\newcommand{\bra}[1]{ {\left\langle #1 \right|} }

\usepackage{epsfig,amssymb,graphics}
\usepackage{hyperref}
\usepackage{calc}
\usepackage{amsmath}
\usepackage{bm}
\pdfoutput=1

\begin{document}

\markboth{S. Guha, et. al.}
{Quantum Auctions}

\catchline{}{}{}{}{}

\title{Quantum Auctions using Adiabatic Evolution: \\ The Corrupt Auctioneer and Circuit Implementations}

\author{Saikat Guha}

\address{Network Technologies Group, BBN Technologies, 10 Moulton St., MS 6/3B, \\
Cambridge, MA 02138,
United States of America\\
sguha@bbn.com}

\author{Tad Hogg}

\address{Social Computing Laboratory, Hewlett-Packard Laboratories, 1501 Page Mill Rd., MS 1123, \\
Palo Alto, CA 94304-1100,
United States of America\\
tad.hogg@hp.com}

\author{David Fattal}

\address{Information and Quantum Systems Laboratory, Hewlett-Packard Laboratories, 1501 Page Mill Rd., MS 1123, \\
Palo Alto, CA 94304-1100,
United States of America\\
david.fattal@hp.com}

\author{Timothy Spiller}

\address{Information and Quantum Systems Laboratory, Hewlett-Packard Laboratories, Filton Road, Stoke Gifford, \\
Bristol BS34 8QZ, 
United Kingdom \\
timothy.spiller@hp.com}

\author{Raymond G. Beausoleil}

\address{Information and Quantum Systems Laboratory, Hewlett-Packard Laboratories, 1501 Page Mill Rd., MS 1123, \\
Palo Alto, CA 94304-1100,
United States of America\\
ray.beausoleil@hp.com}

\maketitle


\begin{abstract}
We examine a proposed auction algorithm using quantum states to represent bids and distributed adiabatic search to find the winner \cite{hogg07}. When the auctioneer follows the protocol, the final measurement giving the outcome of the auction also destroys the bid states, thereby preserving privacy of losing bidders. We describe how a dishonest auctioneer could alter the protocol to violate this privacy guarantee, and present methods by which bidders can counter such attacks. We also suggest possible quantum circuit implementations of the auction protocol, and quantum circuits to perpetrate and to counter attacks by a dishonest auctioneer.
\end{abstract}


\section{Introduction}

The advent of quantum computing has brought home some remarkable, though a handful of, algorithms with varying degrees of computational advantages over their classical counterparts~\cite{shor94,grover96}  -- viz. Shor's factoring algorithm, Grover's search algorithm, quantum simulation, etc.. These algorithms harp on the quantum resources of entanglement and superposition to gain their computational power. In order to see real computational advantage using these algorithms over their classical counterparts in the foreseeable future, one would need to maintain quantum coherence of many qubits for much longer times than is possible now or likely to be realized in the near future.

In \cite{hogg07}, Hogg et. al. pick instead an economic application -- auctions, and show that even using a few qubits, there are many interesting advantages that one may obtain using a quantum algorithm, over existing classical approaches to solve the auctions problem. The auction method uses quantum superpositions to represent bids and adiabatic quantum search to identify the winning bid. An auctioneer performs the final measurement of the quantum state obtaining a unique outcome for the auction while simultaneously destroying the bid states, thereby not learning the bid values of any of the losing bidders,  and hence preserving their privacy. By introducing entanglement between bid states, participants can arrange for correlations among their bids with the assurance that this entanglement will not be observable by others. The method applies to a variety of auction types, e.g., first or second price, and for auctions involving either a single item or bundled items (combinatorial auctions).

In \cite{hogg07}, the authors examine some ways bidders could cheat, and some ways to counter such attacks by bidders. One such example that was examined was where dishonest bidders could create a non-zero probability of a non-maximum bid to win the auctions, by preparing a non-uniform superposition of their bidding preferences, as opposed to a uniform superposition as prescribed by the protocol. In this paper, we analyze various ways in which a corrupt auctioneer can try to tweak the protocol to learn the bid values of the losing bidders without being caught, and at the same time obtaining the correct outcome of the auction. We describe various ways by which such attacks can be implemented, and ways by which suspicious bidders may arrange to counter such attacks perpetrated by the corrupt auctioneer. We study the behavior of the auctions protocol in the simple case of an auction for a single item, and show how these attacks and counter-attacks work, and how they affect the convergence time of the adiabatic search. We also describe quantum circuit implementations of the auction algorithm.

The paper is structured in the following manner. Section 2 overviews the auctions protocol introduced in \cite{hogg07} and establishes the notation. In Section 3, we analyze the performance of the protocol for a simple auction example with two bidders and one item. Section 4 describes quantum circuit implementations of the auctions protocol. In Section 5, we discuss the possible attacks by the corrupt auctioneer and respective counter-attacks by bidders, and circuit implementations thereof. Section 6 summarizes the paper and discusses possibilities for future work.

\section{Quantum Auctions by Adiabatic Search Protocol}
\label{adiabaticprotocol}

In this section we describe a quantum auctions protocol that takes advantage
of the compact and expressive bidding language of quantum
superpositions. The quantum auction differs from its classical
counterpart in two main ways:
\begin{enumerate}
\item The bids are encoded using qubits, instead of bits - which allows for quantum superpositions
\item A quantum search algorithm is used to find the best allocation of items to the bidders.
\end{enumerate}

In our auctions protocol, each bidder selects an operator that
produces the desired bid from a pre-specified initial state (e.g.,
all bits set to zero). The auctioneer repeatedly asks the bidders to
apply their individual operators in a distributed implementation of
a quantum search to determine the winning bid. We use the adiabatic
search method~\cite{farhi01} since it applies directly to finding
optimal values and allows improving incentive compatibility with
simple design modifications~\cite{hogg07}. It is thus better suited to auctions
than decision-problem searches such as Grover's unstructured search
method~\cite{grover96}.

Let us consider $m$ bidders $B_1, B_2, \ldots B_m$ bidding on $n$ items $I_1, I_2, \ldots, I_n$. There is an auctioneer who distributes $p$ qubits to each bidder, i.e. a total of $mp$ qubits, all initialized in their $|0\rangle$ states, viz. $|\Psi_{\rm init}\rangle \equiv |{\psi_{\rm init}}\rangle^{\otimes{m}} = |00{\ldots}0\rangle^{\otimes{m}}$. Let us run through a step-by-step development of our setup of the auctions problem and a solution using quantum adiabatic search algorithm ---

\begin{enumerate}

\item {\bf {Allocation}}: Let us define an ``allocation" $|x\rangle$ as a $mp$-qubit `basis-state' (i.e. a $2^{(mp)} \times 1$ column vector). A `basis-state' in the computational basis is defined to be a product-state, that is a product of $|0\rangle$'s and $|1\rangle$'s.  The allocation given by

\begin{equation}
|x\rangle = |I_{i_1},b_{i_1}^{(1)}; \cdots; I_{i_m},b_{i_m}^{(m)}\rangle
\label{eq:allocation}
\end{equation}
is interpreted by the auctioneer as --- {\em {``Assign item $I_{i_k}$ to bidder $B_k$ and charge him $\$b_{i_k}^{(k)}$"}}, for each $i_k \in \left\{1,2,\cdots,n\right\}$. A couple of points to note here are as follows:
\begin{itemize}
\item In an allocation $|x\rangle$, each of the $m$ `item-bid' pairs is a $p$-qubit `basis-state' out of which $r = \floor{\log{n}}+1$ qubits are needed to represent the `item' (as there are $n$ items), and the remaining $q = p - r$ qubits are available to represent the dollar-value of the bid up to a `resolution' of $2^q$ possible numbers.
\item An allocation $|x\rangle$ is called {\bf {infeasible}} if and only if $|x\rangle$ assigns one item to more than one bidder, or assigns no items to any bidder. A state that is not infeasible, will be termed  {\bf {feasible}}.
\item A bid value of $b_{i_k}^{(k)} = 0$ will be interpreted as ``Don't assign item $I_{i_k}$ to bidder $B_k$."
\item An allocation $|x\rangle$ is called {\bf {plausible}} if and only if for all $k \in \left\{1,2,\cdots,m\right\}$, $\$b_{i_k}^{(k)}$ is a bid-value that $B_k$ is willing to pay for item $I_{i_k}$ taking into consideration all possible `correlations' $B_k$ might have with other people's bid-values for all other items that appear in $|x\rangle$.
\item Define a ``payoff" function $F(.)$ for the auctioneer, so that $F(x)$ is the ``value" that the auctioneer earns if he were to choose to announce allocation $|x\rangle$. $F(x) = 0$, if $|x\rangle$ is an {\bf {infeasible}} allocation. For a standard `first-price' auction $F(x) = \sum_{k=1}^mb_{i_k}^{(k)}$ is the total revenue the auctioneer earns if a feasible allocation $|x\rangle$ (as defined above) is carried out.
\item The $p$-qubit null state is defined as $|\phi\rangle \equiv |00\cdots{0}\rangle$.
\end{itemize}

\item {\bf {Goal}}: The goal of setting up this problem is to first have the bidders collectively create a uniform superposition of all possible {\bf {plausible}} allocations (which possibly might consist of some infeasible states). Now, we want to setup the adiabatic quantum search algorithm, so that the search starts out with this uniform superposition of all plausible states, and through several iterations of operations on $mp$ qubits the state slowly evolves into the {\bf {plausible}} state in that superposition that has the maximum payoff $F(x)$, after which a measurement of all the qubits reveals to the auctioneer the payoff-maximizing allocation $|x^*\rangle$ with very high probability.

\item {\bf {Creating the uniform superposition}}: Because of the distributed nature of this search, one way the bidders can collectively create a uniform superposition of all plausible states, is if each bidder $B_j$ creates a $p$-qubit {\bf {bidding-state}} $\ket{\psi_j}$ -- a state containing a uniform superposition of all their bid preferences.

\begin{equation}
|\psi_j\rangle = \ket{\phi} + \sum_{i \in {\cal {I}}_j} \ket{I_{i}^{(j)},b_{i}^{(j)}}
\end{equation}
where ${\cal {I}}_j \subseteq \left\{1,2,\cdots,n\right\}$ is the set of indices representing the items bidder $B_j$ is interested in bidding on. Each bidder is expected to come up with a $2^p \times 2^p$ unitary evolution matrix $U_i$, that creates their respective bidding-state $|\psi_j\rangle$ from the initialized $p$-qubit state\footnote{note that the condition imposed by Eq.~\eqref{eq:Uj} on the choice of the bidding operator $U_j$ just specifies the first column of the $2^p \times 2^p$ unitary matrix $U_j$. It is obvious that there are many ways the bidders can extend the first columns of their own bidding operators $U_j$ into a complete orthonormal (CON) basis of column vectors to construct the full bidding operator $U_j$. We later show in this paper the adiabatic search algorithm that we describe works as desired when the bidders create their bidding operators using a Hadamard-like construction.\label{footUj}} $|\psi_{\rm init}\rangle = |00\cdots{0}\rangle$:

\begin{equation}
U_j|\psi_{\rm init}\rangle = |\psi_j\rangle
\label{eq:Uj}
\end{equation}

If two bidders $B_j$ and $B_k$ wish to correlate their bidding preferences, they would create a joint unitary operator $U_{j,k}$ (a $2^{2p} \times 2^{2p}$ matrix), such that $U_{j,k}\ket{\psi_{\rm init}}^{\otimes 2} = \ket{\psi_{j,k}} \ne \ket{\psi_j} \otimes \ket{\psi_k}$, where $\ket{\psi_{j,k}}$ is the joint bidding state of bidders $B_j$ and $B_k$. Each bidder returns their bidding-states to the auctioneer.  The state of the $mp$ qubits that the auctioneer now has is a uniform superposition of all plausible states. Assuming no bid-correlations for notational simplicity, we have

\begin{eqnarray}
|\Psi_0\rangle &=& U|\Psi_{\rm init}\rangle \nonumber \\
&=& \left(U_1 \otimes \cdots \otimes U_m\right)|\psi_{\rm init}\rangle^{\otimes m} \nonumber \\
&=& \bigotimes_{j=1}^m{\ket{\psi_j}} \nonumber \\
&=& \sum_{x: {\rm plausible}}|x\rangle
\end{eqnarray}
where each term $\ket{x}$ in the above sum is a plausible allocation. Note that there could be several infeasible allocations $\ket{x}$ in the sum.

\item {\bf {Adiabatic search}}: The adiabatic theorem can be used to design various quantum search algorithms. If a system is created in the ground state of a `beginning' Hamiltonian $H_b$, and is made to evolve under the interaction of a slowly changing Hamiltonian, such as

\begin{equation}
H(f) = (1-f)H_b + fH_p, {\qquad} f \in [0,1];
\label{eq:H_interp}
\end{equation}
then provided that none of the higher eigenvalues of $H(f)$ ever intersect with the eigenvalue of the ground state of $H(f)$ in $f \in [0,1]$ and provided the evolution is done `slowly' enough, the system ends up in the ground state of the `problem' Hamiltonian $H_p$, which is (by construction) the solution to our search problem.

We construct a $2^{mp} \times 2^{mp}$ diagonal matrix $H_p$ whose diagonal entries are negatives of the payoff values $F(x)$ for all $2^{mp}$ possible allocations $\ket{x}$ in the computational basis (See Section III for an explicit example). The ground state of $H_p$ is thus the allocation vector $\ket{x_0}$ corresponding to the lowest eigenvalue, i.e. highest payoff $F(x_0)$.

We also construct another $2^{mp} \times 2^{mp}$ diagonal matrix $W$ whose diagonal entries $d(x)$ are say, the Hamming weights of all the $2^{mp}$ possible allocations $\ket{x}$ in the computational basis. The ground state of $W$ is thus the `all-zero' state $|\Psi_{\rm init}\rangle$ with eigenvalue $0$. We define the beginning Hamiltonian of the search as
$$
H_b \equiv UWU^\dagger.
$$
The ground state of $H_b$ is thus $U\ket{\Psi_{\rm init}} = |\Psi_0\rangle = \bigotimes_{j=1}^m\ket{\psi_j} = \sum_{x: {\rm plausible}}|x\rangle$. This is precisely the state in which we start out our adiabatic search. Now, consider the following iterations of a discrete version of the adiabatic search process:
\begin{eqnarray}
\ket{\Psi_s} &=& e^{-i\Delta{H(f)}}\ket{\Psi_{s-1}} \label{eq:iterations_exact} \\
&=& e^{-i\Delta{\left((1-f)H_b+fH_p\right)}}\ket{\Psi_{s-1}} \nonumber \\
&=& e^{-i\Delta{\left((1-f)UWU^{\dagger}+fH_p\right)}}\ket{\Psi_{s-1}} \nonumber \\
&\approx& e^{-i\Delta{(1-f)}UWU^\dagger}e^{-i\Delta{fH_p}}\ket{\Psi_{s-1}} \nonumber \\
&=& U\left(e^{-i\Delta{(1-f)W}}\right)U^{\dagger}e^{-i{\Delta}fH_p}\ket{\Psi_{s-1}} \nonumber \\
&=& UD(\Delta, 1-f)U^{\dagger}P(\Delta, f)\ket{\Psi_{s-1}}
\label {eq:iterations}
\end{eqnarray}
where $P(\Delta, f) \equiv e^{-i{\Delta}fH_p}$, $D(\Delta, f) \equiv e^{-i{\Delta}fW}$, and $\Delta$ is a small discrete parameter which represents a small time interval, over which $H(f)$ can be considered approximately constant. Total number of iterations is $S$, and $f = s/S$, for $s = {1,2,\cdots,S}$. As each bidder's $U_j$ preferentially `rotates' $\ket{\psi_{\rm init}}$ to a $|{\cal {I}}_j|+1$ dimensional subspace of the $2^p$ dimensional space of their $p$ qubits, $U = \bigotimes_{j=1}^mU_j$ `rotates' $\ket{\Psi_{\rm init}}$ to a $\prod_{j=1}^m{\left(|{\cal {I}}_j|+1\right)}$ dimensional subspace of the full $2^{mp}$ dimensional Hilbert space of the $mp$ qubits. For a special Hadamard-like construction of each $U_j$ (see Appendix A), the above iterative search procedure never goes out of the subspace spanned by the allocation vectors in the initial state $\ket{\Psi_0} = \sum_{x: {\rm plausible}}|x\rangle$. The search thus converges towards the allocation vector $\ket{x^*}$ in the initial superposition that has the maximum payoff with probability of success $P_{\rm success} \rightarrow 1$ --- the solution to our problem \ref{footConvergence}! The auctioneer then measures the final state, and announces the winners and winning-bids for each item. 

\item {\bf {Properties}}: Some final comments and properties of the quantum adiabatic search applied to auctions ---

\begin{itemize}
\item The `time-step' $\Delta$ must be chosen in a way, such that the continuous adiabatic limit $T = S{\Delta} \longrightarrow \infty$ is achieved as $S \longrightarrow \infty$. One way to ensure that is to choose\footnote{Given that the adiabatic search is run with a sufficiently small $\Delta$, and a sufficiently large number of iterations $S$, the algorithm is indeed guaranteed to succeed -- and by ``succeed", we mean that the probability of the (honest) auctioneer getting the correct result from his final measurements can be made exceedingly close to $1$. Note that the process never succeeds with probability $1$. As we state in the paper, the `time-step' $\Delta$ must be chosen in a way, such that the continuous adiabatic limit $T = S{\Delta} \longrightarrow \infty$ is achieved as $S \longrightarrow \infty$. One way to ensure that is to choose $\Delta = 1/\sqrt{S}$. In this limit, the probability of success $P_{\rm success} \rightarrow 1$.\label{footConvergence}} $\Delta = 1/\sqrt{S}$.
\item After the search converges, when the auctioneer makes a measurement of all the $mp$ qubits to find the winning allocation $\ket{x^*}$, all information about the bid values of all losing bidders for each item is instantly destroyed. Thus, the quantum auctions protocol protects the privacy of losing bidders.
\item Eq.~\ref{eq:iterations} uses the approximation $e^{(A+B)\Delta} \approx e^{A\Delta}e^{B\Delta}$. A better approximation $e^{(A+B)\Delta} \approx e^{A\Delta/2}e^{B\Delta} e^{A\Delta/2}$ can be used to obtain the following second-order version of the adiabatic iterations for the auctions protocol:
\begin{equation}
\ket{\Psi_s} = UD(\frac{\Delta}{2}, 1-f)U^{\dagger}P(\Delta, f)UD(\frac{\Delta}{2}, 1-f)U^{\dagger}\ket{\Psi_{s-1}}
\label{eq:iterations_FO}
\end{equation}
\item This protocol assumes that the bidders and the auctioneer do their respective parts throughout the running of the protocol honestly, as prescribed by the above protocol. Further additions and modifications in the protocol will be necessary to make it robust to corrupt auctioneers and dishonest bidders who might try to tweak the protocol to cater to their respective motives.
\end{itemize}
\end{enumerate}

\section{A Toy Example}
\label{example}

In this section, we will look at a very simple four-qubit example to illustrate the theory developed above to implement quantum auctions, and will examine convergence behavior of the quantum adiabatic algorithm for this case. We will find it useful later to refer back to this example to illustrate various aspects of quantum auctions using adiabatic search. Let us consider a simple example with one item being auctioned off $(n=1)$ amongst two bidders $B_1$ and $B_2$ $(m=2)$. Let us say that the allowed set of price-values for the item are -- $\$1$, $\$2$, and $\$3$. Lets say that the {\em{bidding states}} corresponding to these three bid values are

\begin{eqnarray}
|\psi_1\rangle = \frac{|00\rangle + |01\rangle}{\sqrt{2}}  \nonumber \\
|\psi_2\rangle = \frac{|00\rangle + |10\rangle}{\sqrt{2}}  \nonumber \\
|\psi_3\rangle = \frac{|00\rangle + |11\rangle}{\sqrt{2}}
\label{eq:biddingstates}
\end{eqnarray}

\begin{figure}
\begin{center}
\includegraphics[height=6cm,width=8cm,angle=0]{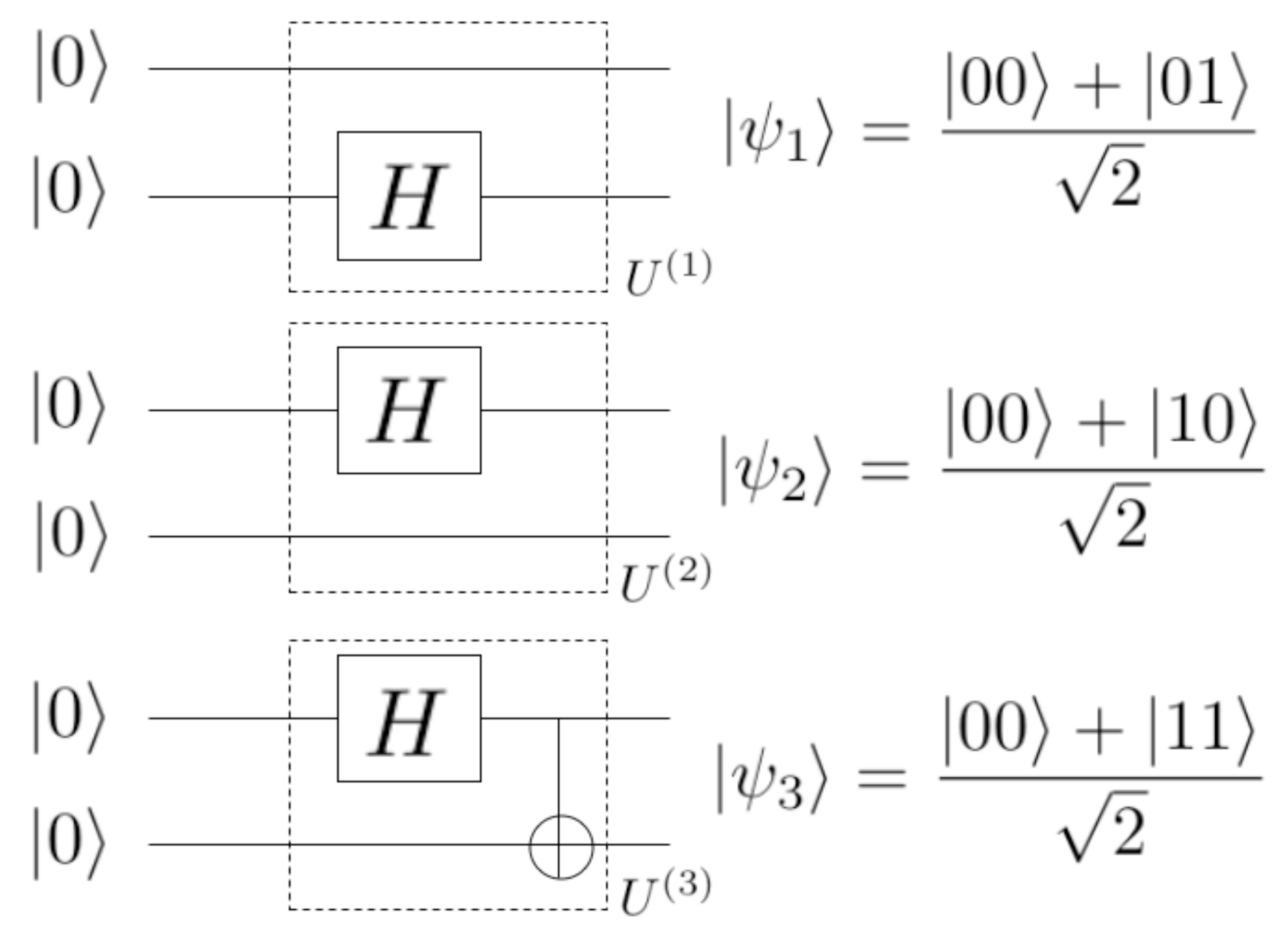}
\end{center}
\caption{Two-qubit unitary operators $U^{(i)}$ that the bidders choose from, in order to generate their `bidding states'. $U^{(i)}|00\rangle = |\psi_i\rangle$, where $|\psi_i\rangle$ is the `bidding state' corresponding to the bid value of $\$i$.}
\label{fig:unitops}
\end{figure}

The bidders choose their unitary operators $U_1$ and $U_2$ out of the set of three unitary operators $\left\{U^{(1)}, U^{(2)}, U^{(3)}\right\}$ in order to generate their corresponding bidding states by applying the operator on the two-qubit initialized state $|\psi_{\rm init}\rangle = |00\rangle$, viz. $U^{(i)}|00\rangle = |\psi_i\rangle$. The unitary operators can be implemented by Hadamard and Controlled-not (CNOT) gates as shown in figure~\ref{fig:unitops}\footnote{See Appendix A, for a discussion on why such a Hadamard-like construction guarantees that the condition that the adiabatic search described in Eq.~\eqref{eq:iterations} always remains in the initial subspace of the bidding states, $|\Psi_0\rangle$}. The final Hamiltonian for the adiabatic search $H_p$ is a $16 \times 16$ diagonal matrix, the diagonal elements of which are negatives of the payoff values for the auctioneer, $F(x)$ for each of the $16$ possible `allocations' $|x\rangle$ represented as $16 \times 1$ unit-vectors in the standard Kronecker basis. Given the payoff values in Table~\ref{table:payoff}, $H_p = {\rm diag}\left\{0, -1, -2, -3, -1, 0, 0, 0, -2, 0, 0, 0, -3, 0, 0, 0\right\}$. Let us define the initial Hamiltonian $H_i \equiv UWU^\dagger$, where $W$ is a $16 \times 16$ diagonal matrix, that assigns a $0$ eigenvalue to the initial state: $|\psi_{\rm init}\rangle \otimes |\psi_{\rm init}\rangle = |00\rangle \otimes |00\rangle$, and higher eigenvalues to all other states. One possibility is to define a diagonal matrix $W$ with the diagonal elements being the Hamming weights of the state-vectors, in which case $W = {\rm diag}{\left\{0, 1, 1, 2, 1, 2, 2, 3, 1, 2, 2, 3, 2, 3, 3, 4\right\}}$ in the standard Kronecker basis.

\begin{figure}
\begin{center}
\includegraphics[height=7cm,width=9cm,angle=0]{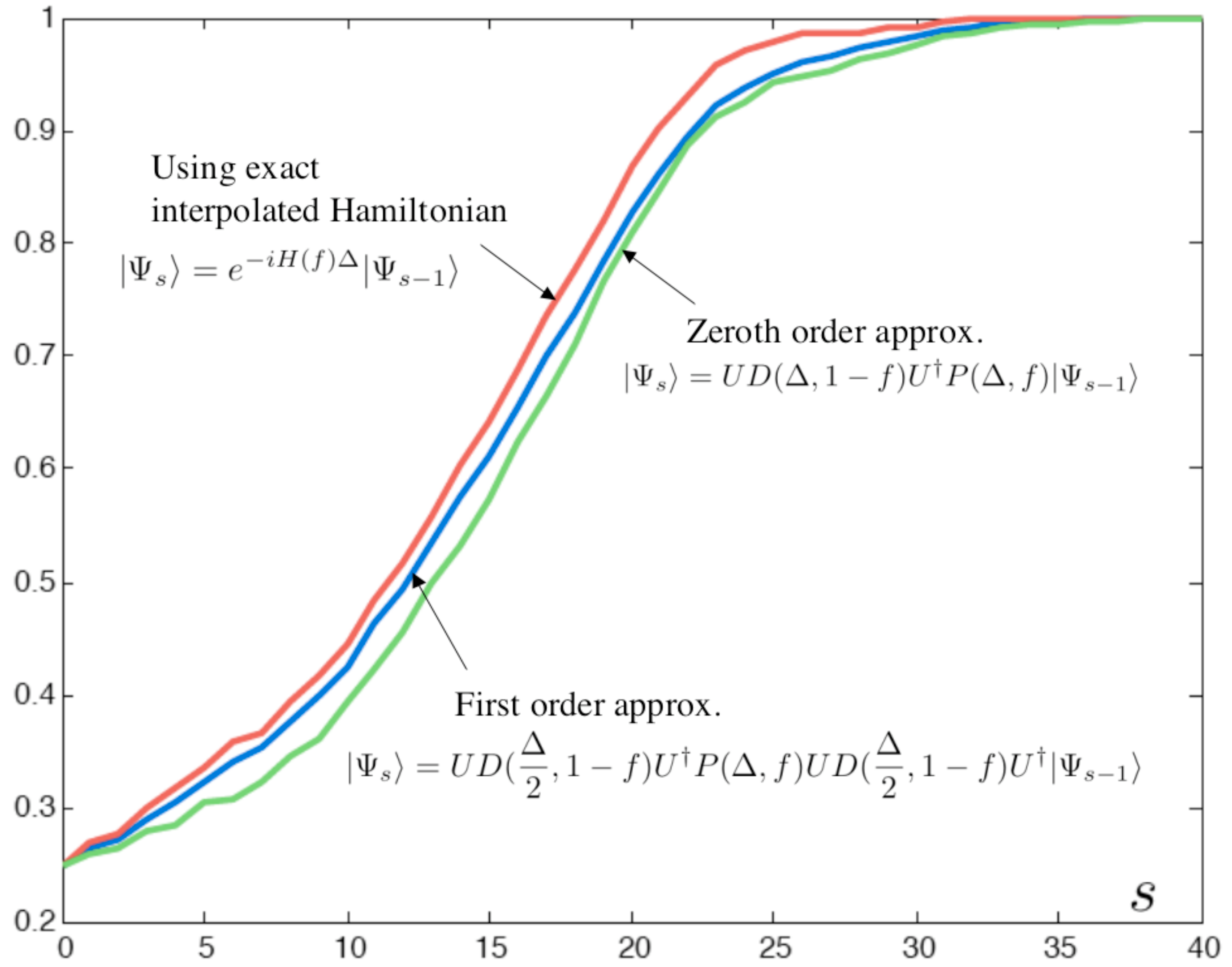}
\end{center}
\caption{This figure compares the convergence rates of the adiabatic search algorithm applied to the auctions problem for a simple case of two bidders bidding on a single item, for three cases --- (a) when the exact expression for the interpolated Hamiltonian (Eq.~\ref{eq:H_interp}) is used, (b) when a zeroth order approximation (Eq.~\ref{eq:iterations}) is used, and (c) for a first order approximation (Eq.~\ref{eq:iterations_FO}). Each bidder has two qubits to express their bid values and the price of the item can only take the values $\$1$, $\$2$, and $\$3$. In all the above plots, the bidders' price values are $\$2$ and $\$3$ respectively, and $S=40$, $\Delta = 1$. The x-axis is $s \in [0,S]$ is the discrete index that interpolates between the Hamiltonians $H_i$ and $H_p$. The y-axis is the probability (as a function of $s$) that the auctioneer will pick the maximum bidder as the winner if he makes a measurement on $|\Psi_s\rangle$.}
\label{fig:convergence_methods}
\end{figure}

\begin{figure}
\begin{center}
\includegraphics[height=7cm,width=9cm,angle=0]{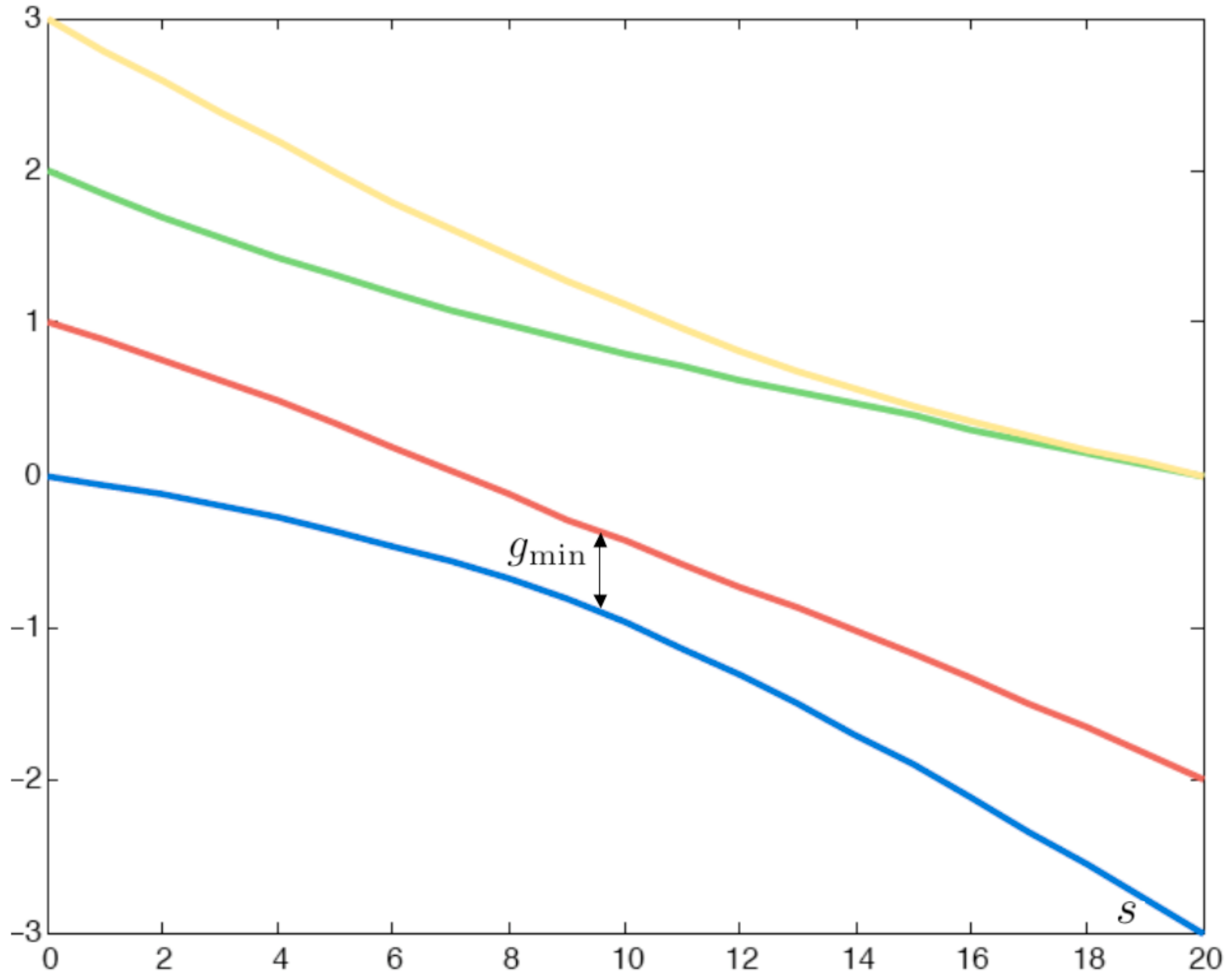}
\end{center}
\caption{This figure shows the four eigenvalues of the interpolated Hamiltonian $H(f)$ as a function of the number of iterations $s$. $f = s/S$ for $s=1,2,\cdots,S$, where $S=20$ is the total number of iterations. It also shows the minimum gap $g_{\rm min}$ between the energy-values of the ground-state and the first excited state of $H(f)$. We plot here the four eigenvalues of $H(f)$ for the particular case when the bidders bid $\$2$, and $\$3$ on the item. The reason the eigenvalues do not cross is that the other eigenstates are always uncoupled from these four, as the search never leaves the initial subspace\ref{footUj}. We can restrict our attention to the particular $4-by-4$ subspace of the original $16-by-16$ matrix space -- for a given choice of a pair of bid values.}
\label{fig:eigenvalues}
\end{figure}

\begin{figure}
\begin{center}
\includegraphics[height=7cm,width=9cm,angle=0]{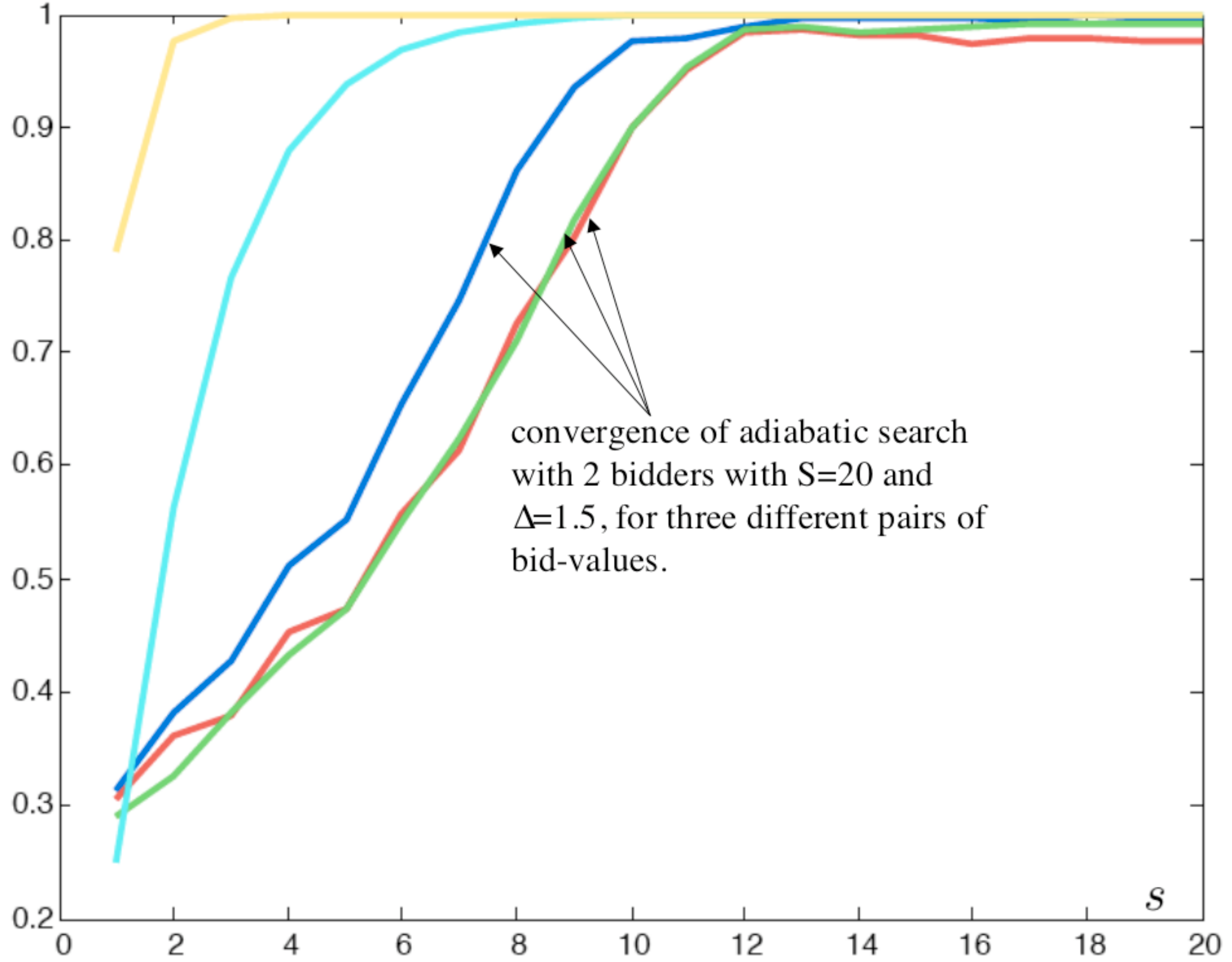}
\end{center}
\caption{This figure shows the convergence of the adiabatic search algorithm applied to the auctions problem for a simple case of two bidders bidding on a single item whose price can only take the values $\$1$, $\$2$, and $\$3$. Each bidder has two qubits to express their bid values. The solid red, green and the dark-blue lines show the probability that the auctioneer will measure the winning-state if he makes a measurement at the $f^{\rm th}$ step of the adiabatic process, where $f=s/S$, for a total of $S=20$ steps and $\Delta = 1.5$. The three lines correspond to the convergence behavior for pairs of bid values $\left\{2,3\right\}$, $\left\{1,2\right\}$, and $\left\{1,3\right\}$ respectively. The cyan line plots the probability that a corrupt auctioneer will learn the bid-values of all bidders correctly, as a function of $f$, if he uses $|00\rangle$-probe-states and a qubit-by-qubit measurement in the computational basis. The yellow line plots the `learning curve' of the corrupt auctioneer who uses two-qubit optimum joint measurement (POVM) in order to learn the bid-values.}
\label{fig:convergence}
\end{figure}

In Figure~\ref{fig:convergence_methods}, we compare the performance of the three approximations to the adiabatic algorithm --- (a) when the exact expression for the interpolated Hamiltonian (Eq.~\ref{eq:iterations_exact}) is used, (b) when a zeroth order approximation (Eq.~\ref{eq:iterations}) is used, and (c) for a first order approximation (Eq.~\ref{eq:iterations_FO}). They are all discrete-time approximations of the continuous-time adiabatic algorithm, with discretization parameters $S = 40$ and $\Delta = 1$. We find that the first-order approximation converges faster than the zeroth order approximation. In a later section, we will find that we encounter very little overhead in implementing the first-order-approximate version of the algorithm, as compared to the zeroth-order-approximate version. In Figure~\ref{fig:convergence}, we plot the convergence rates for the above two-bidder example, for all three possible pairs of possible bid-values --- $\left\{2,3\right\}$, $\left\{1,2\right\}$, and $\left\{1,3\right\}$ respectively. We used $S = 20$ and $\Delta = 1.5$ for these plots. It shows that the algorithm converges to the revenue-maximizing solution irrespective of the actual bids placed. The convergence speed of adiabatic search scales inversely with the minimum energy-gap $g_{\rm min}$ between the ground-state and the first-excited-state eigenvalues of the interpolated Hamiltonian $H(f)$ (see Eq.~\ref{eq:H_interp}). Figure~\ref{fig:eigenvalues} shows the four eigenvalues as a function of the number of iterations $s$, when the bid values are $\$2$ and $\$3$. Its also worth noting that no matter how high a value of $S\Delta$ one might choose, there is still a finite (even though vanishingly small, as $S\Delta \to \infty$) probability that the quantum auctions protocol will not return the optimal value (i.e., highest bid) as the final outcome.

\begin{table}
\tbl{Auction payoffs for two players $B_1$ and $B_2$ bidding on one
item. Each player has two qubits. The `allocation' $\ket{x} = \ket{q_1
q_2}_{B_1} \ket{q_3 q_4}_{B_2}$ has a payoff given by $F(x) =  -\bra{x}
H_p\ket{x}$, where $H_p$ is a diagonal matrix that constitutes the final Hamiltonian of the adiabatic search. A \emph{desirable} final state consists of only
one player with a nonzero bid (i.e., the winner of the auction.) In
other words, $F(x) \neq 0$ only if $\ket{\psi}_{B_1} = \ket{00}_{B_1}$
\emph{or} $\ket{\psi}_{B_2} = \ket{00}_{B_2}$, but not both.}
{\begin{tabular}{|ccc|}
\hline
$\ket{q_1 q_2}_{B_1}$ & $\ket{q_3 q_4}_{B_2}$ & $F(x)$ \\
\hline
 00 &  00 &  0  \\
 00 &  01 &  1  \\
 00 &  10 &  2  \\
 00 &  11 &  3  \\
 01 &  00 &  1  \\
 01 &  01 &  0  \\
 01 &  10 &  0  \\
 01 &  11 &  0  \\
 10 &  00 &  2  \\
 10 &  01 &  0  \\
 10 &  10 &  0  \\
 10 &  11 &  0  \\
 11 &  00 &  3  \\
 11 &  01 &  0  \\
 11 &  10 &  0  \\
 11 &  11 &  0  \\
\hline
\end{tabular}}
\label{table:payoff}
\end{table}

\section{Quantum Circuit Implementations}
\label{implementation}

We will first begin with deriving approximate circuit-level implementations of the quantum auctions protocol for the simple $4$-qubit example described in the previous section. We will also sketch a scheme to implement the auctions protocol in general. Let us split our discussion into three parts --- (1) Bidders' unitary operators $U_i$'s, (2) the incremental initial Hamiltonian $D(\Delta, f)$, and (3) the incremental final Hamiltonian $P(\Delta, f)$.

\begin{enumerate}

\item {\bf Bidders' unitary operators} --- These operators are the easiest to implement using quantum circuits. We know that (in the absence of correlated bidding preferences) the joint unitary operator $U$ in Eq.~\ref{eq:iterations} and Eq.~\ref{eq:iterations_FO} is a tensor product of the bidders' individual unitary operators $U_i$, $i=1,2,\cdots,m$. Each unitary $U_i$ is a $p$-qubit operator which essentially performs the 2-qubit ``Hadamard" operation in some $2$-dimensional subspace of the Hilbert space of $p$ qubits. To be more specific, suppose we want to implement the unitary $U_i$ that takes the $p$-qubit state $|00\cdots{0}\rangle$ to the $p$-qubit state $|\psi_i\rangle = |00101\cdots{01}\rangle$ where lets say $t$ qubits with indices $i_k \in \left\{1,2,\cdots,p\right\}$ in $|\psi_i\rangle$ are $|1\rangle$ and the rest are $|0\rangle$. Lets define $i_{\rm min} = {\min_{k=1}^t}\left\{i_k\right\}$. One possible way to implement $U_i$ is --- Apply a Hadamard operation $H$ to qubit $i_{\rm min}$, followed by $(t-1)$ C-NOT operations with the $i_{\rm min}^{\rm th}$ qubit as the control-qubit and the remaining $t-1$ qubits $i_k$ as the target qubits. For concrete examples, see Figure~\ref{fig:implement_U} and Figure~\ref{fig:unitops}.

\begin{figure}
\begin{center}
\includegraphics[height=3.5cm,width=10cm,angle=0]{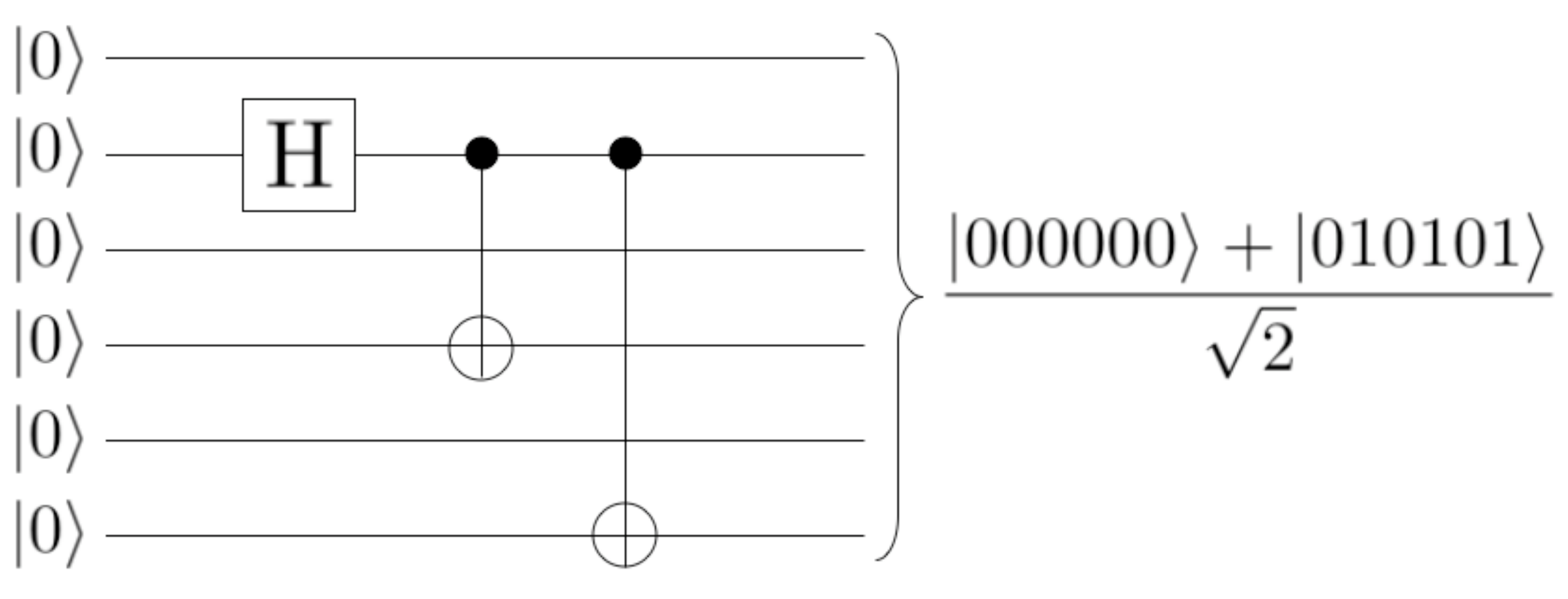}
\end{center}
\caption{This figure shows the quantum circuit to implement the unitary operator $U$ for a bidder to create a $p$-qubit bidding state $(p=6)$ starting from initialized state $|\psi_{\rm init}\rangle = |0\rangle^{\otimes{6}}$. In this example, $U|0\rangle^{{\otimes}6} = |010101\rangle$.}
\label{fig:implement_U}
\end{figure}

\item {\bf Incremental initial Hamiltonian $D(\Delta, f)$} --- This incremental Hamiltonian is used by the auctioneer once in each iteration of the protocol. From the discussion in section \ref{example}, we know that one way to realize the initial Hamiltonian for the search problem is to have
$$
D(\Delta, f) = e^{-i\Delta{fW}},
$$
where, $W$ is a diagonal matrix with the diagonal elements being the Hamming weights of the $2^{mp}$ $mp$-qubit standard basis-vectors. In the case of the $4$-qubit example in section \ref{example}, $W = {\rm diag}{\left\{0, 1, 1, 2, 1, 2, 2, 3, 1, 2, 2, 3, 2, 3, 3, 4\right\}}$ in the standard Kronecker basis. Consider the following matrix identity:

\begin{eqnarray}
\left(\begin{array}{cc}
1 & 0 \\
0 & e^{-i\delta}
\end{array}
\right) \otimes \left(\begin{array}{cc}
1 & 0 \\
0 & e^{-i\delta}
\end{array}
\right) &=& \left(\begin{array}{cccc}
1 & 0 & 0 & 0 \\
0 & e^{-i\delta} & 0 & 0 \\
0 & 0 & e^{-i\delta} & 0 \\
0 & 0 & 0 & e^{-i2\delta}
\end{array}
\right) \nonumber \\
&=& \exp({-i\delta\left( \begin{array}{cccc}
0 & 0 & 0 & 0 \\
0 & 1 & 0 & 0 \\
0 & 0 & 1 & 0 \\
0 & 0 & 0 & 2
\end{array}\right)}) \nonumber
\end{eqnarray}

Extending this idea, it is easy to see that

\begin{equation}
D(\Delta, f) = e^{-i\Delta{fW}} = \left(\begin{array}{cc} 1 & 0 \\ 0 & e^{-if\Delta}\end{array}\right)^{{\otimes}4}
\end{equation}

This method generalizes to $mp > 4$ qubits in an analogous fashion. Its worth mentioning at this point that it just takes $mp = 4$ independent single-qubit rotations to all the qubits, to implement $D(\Delta, f)$, where $f$ gets incremented by $1/S$ in each subsequent iteration of the protocol.

\item {\bf Incremental final Hamiltonian $P(\Delta, f)$} --- Let us consider the following arbitrary payoff table (table~\ref{table:payoff_general}) where there are $N$ qubits in total $i_1, i_2, \cdots, i_N$. The last column $P({\underline{i}})$ lists the payoff values for the $2^N$ $N$-qubit basis-states $|i_1, i_2, \cdots, i_N\rangle$. We want to come up with a $2^N \times 2^N$ diagonal matrix $H_p$ whose diagonal values are the payoff values $-P({\underline{i}})$, i.e. ${\langle}{\underline{i}}|{H_p}|{\underline{j}}\rangle = -P({\underline{i}})\delta_{{\underline{i}}{\underline{j}}}$. It is easy to see that, if $Z_k$ is the Pauli $\sigma_z$ operator acting on the $k^{\rm th}$ qubit, then $H_p$ is given by:

\begin{table}
\tbl{Construction of $H_p$ for arbitrary payoff table}
{\begin{tabular}{|cccc|c|}
\hline
$i_1$ & $i_2$ & $\ldots$ & $i_N$ & $P({\underline{i}})$ \\
\hline
0 & 0 & \ldots & 0 & $p_{00\cdots{0}}$ \\
0 & 0 & \ldots & 1 & $p_{00\cdots{1}}$ \\
. & . & \ldots & . & $\cdots$ \\
1 & 1 & \ldots & 1 & $p_{11\cdots{1}}$ \\
\hline
\end{tabular}}
\label{table:payoff_general}
\end{table}

\begin{equation}
H_p = \frac{1}{2^N}\sum_{{\underline{i}} \in \left\{2^N\right\}}(-P({\underline{i}}))\prod_{k=1}^N\left(1+(-1)^{i_k}Z_k\right)
\end{equation}

Now let us construct the final Hamiltonian $H_p$ for the $4$-qubit example above (see table~\ref{table:payoff}). If we renumber the $4$ qubits $|q_1q_2\rangle_{B_1}|q_3q_4\rangle_{B_2}$ as $1$, $2$, $3$, and $4$ respectively, we are led to the explicit construction:

\begin{eqnarray}
H_p &= - \frac{1}{4} \left(1+Z_{1}\right) \left(1+Z_{2}\right) \left[ 0 \cdot \frac{1}{4} \left(1+Z_{3}\right) \left(1+Z_{4}\right) + 1 \cdot \frac{1}{4} \left(1+Z_{3}\right) \left(1-Z_{4}\right) \right. \nonumber \\
 & \qquad \qquad \qquad \qquad \qquad  + \left. 2 \cdot \frac{1}{4} \left(1-Z_{3}\right) \left(1+Z_{4}\right) + 3 \cdot \frac{1}{4} \left(1-Z_{3}\right) \left(1-Z_{4}\right) \right] \nonumber \\
 &- \frac{1}{4} \left(1+Z_{3}\right) \left(1+Z_{4}\right) \left[ 0 \cdot \frac{1}{4} \left(1+Z_{1}\right) \left(1+Z_{2}\right) + 1 \cdot \frac{1}{4} \left(1+Z_{1}\right) \left(1-Z_{2}\right) \right. \nonumber \\
 & \qquad \qquad \qquad \qquad \qquad  + \left. 2 \cdot \frac{1}{4} \left(1-Z_{1}\right) \left(1+Z_{2}\right) + 3 \cdot \frac{1}{4} \left(1-Z_{1}\right) \left(1-Z_{2}\right)
 \right] \nonumber \\
 &=- \frac{1}{16} \left[\left(1+Z_{1}\right) \left(1+Z_{2}\right)
\left(6 - 4 Z_{3} - 2 Z_{4}\right) + \left(1+Z_{3}\right) \left(1+Z_{4}\right)
\left(6 - 4 Z_{1} - 2 Z_{2}\right)\right] \nonumber \\
 &=-\frac{1}{16}\left[12+2Z_1+4Z_2+2Z_3+4Z_4+6Z_1Z_2-6Z_2Z_3\right. \nonumber \\
 &-8Z_3Z_1-6Z_1Z_4-4Z_2Z_4+6Z_3Z_4 \nonumber \\
 & \qquad \left.{-4Z_1}Z_2Z_3-2Z_2Z_3Z_4-2Z_1Z_2Z_4-4Z_1Z_3Z_4\right].
\label{eq:Hp}
\end{eqnarray}

Now, in order to implement the incremental final-Hamiltonian unitary operator $P(\Delta,f)=e^{-i\Delta{fH_p}}$, let us apply the `zeroth' order approximation (for $\Delta \ll 1$) when exponentiating the expression in Eq.~\ref{eq:Hp}, i.e.

\begin{eqnarray}
P(\Delta,f) &= e^{-i\Delta{fH_p}} \approx e^{i3f\Delta/4}e^{if\Delta{Z_1}/8}e^{if\Delta{Z_2}/4} \times \nonumber \\
& \qquad e^{if\Delta{Z_3}/8}e^{if\Delta{Z_4}/4}e^{i3f\Delta{Z_1Z_2}/8} \times \nonumber \\
& \qquad e^{-i3f\Delta{Z_2Z_3}/8}{\cdots}e^{-if\Delta{Z_1Z_3Z_4}/4}
\label{eq:Pf_approx}
\end{eqnarray}

Using the quantum circuit in Figure~\ref{fig:circuit1}, we can implement any unitary operator of the form $e^{i{Z \otimes Z \otimes \cdots \otimes Z}\Delta}$ using only two-qubit operations (control-NOT and phase gates) as shown in the figure. Hence, a collection of two-qubit gates can be put together to construct the incremental final Hamiltonian $P(\Delta,f)$.

\begin{figure}
\begin{center}
\includegraphics[height=6cm,width=12cm,angle=0]{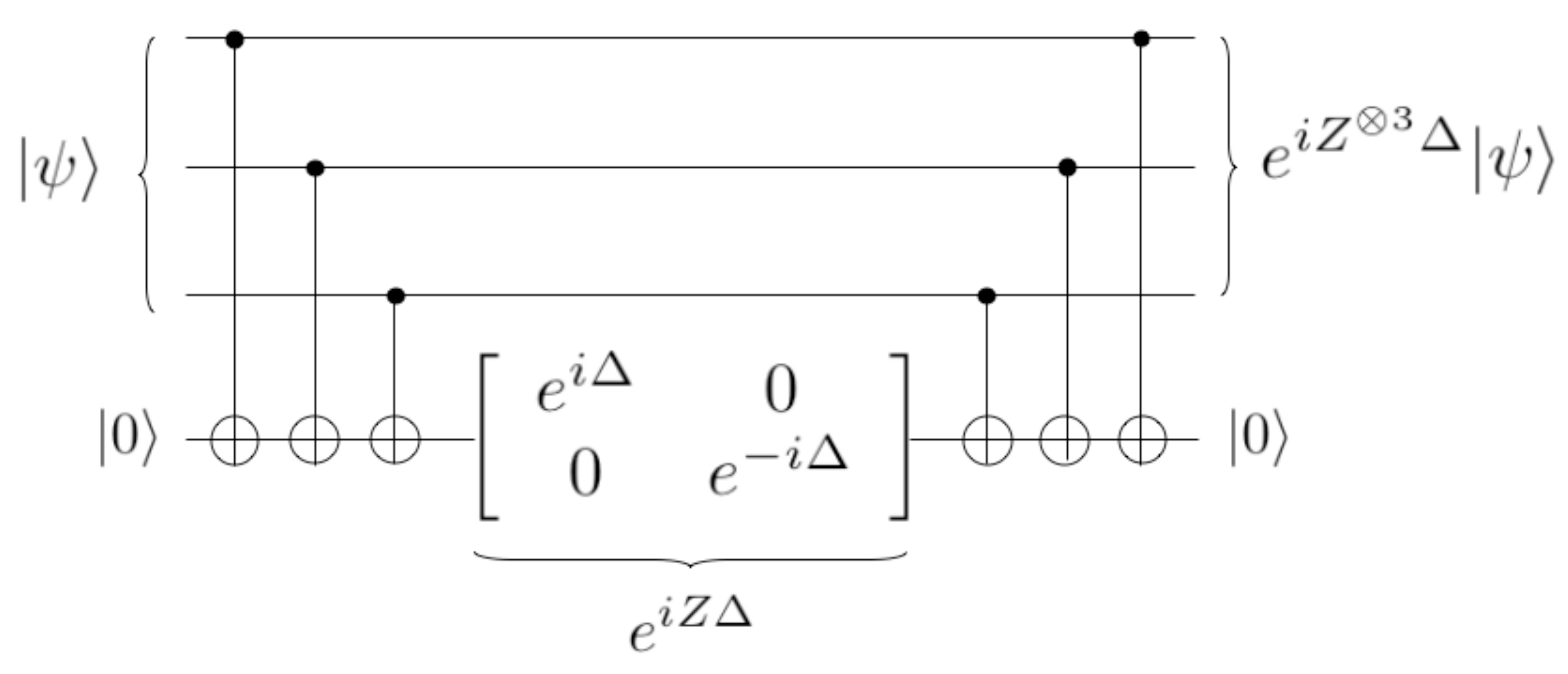}
\end{center}
\caption{This figure shows the quantum circuit to implement the unitary operator $e^{i{Z \otimes Z \otimes  Z}\Delta}$ on an arbitrary $3$-qubit state $|\psi\rangle$. This method can be readily generalized to implementing $e^{i{Z \otimes Z \otimes \cdots \otimes Z}\Delta}$ using C-NOT (Control-NOT) and phase operations, both of which are $2$-qubit unitary operators. Cascade of such circuits can be used to construct the incremental final Hamiltonian $P(\Delta,f)$.}
\label{fig:circuit1}
\end{figure}

\end{enumerate}

\section{The Corrupt Auctioneer}

In this section, we are going to look at ways the security of the quantum auctions protocol can be attacked by a corrupt auctioneer, whose intention is to learn the bid values of all the bidders by deviating from the rules set by the auctions protocol using adiabatic evolution. We will look at two major ways for a corrupt auctioneer to attack the protocol, and will suggest ways to combat the problem. All discussion in this section will pertain to the $4$-qubit toy example elaborated in section \ref{example}, unless stated otherwise. The extension of these ideas to the general case is straightforward.

\subsection{Attack by measurement}

\begin{enumerate}

\item {\bf {The attack mechanism: `probes and measurement'}} --- Imagine that every time the auctioneer gets back the qubits from the bidders once they have applied the joint Hermitian conjugate of their bidding operators $U^\dagger = U_1^\dagger \otimes U_2^\dagger$, instead of operating the qubits with $D(\Delta, 1-f)$, he trashes all the qubits and sends out fresh copies of the initialized state: $|\psi_{\rm probe}\rangle = |00\rangle \otimes |00\rangle$. The bidders, unaware of this, assume that the auctioneer is running the adiabatic evolution as was promised, and apply their unitary bidding operators $U = U_1 \otimes U_2$ on the qubits. Now that the auctioneer has a tensor-product of two fresh copies of the bidding states (Eq.~\ref{eq:biddingstates}), he makes a symbol-by-symbol measurement on each qubit in the computational basis, i.e. using the positive operator-valued measure (POVM) elements $\left\{|0\rangle\langle{0}|,|1\rangle\langle{1}|\right\}$. As each bidding state is an equal superposition of $|00\rangle$ and the `price state' of the bidders, the auctioneer learns both the bidder's bid-values with probability $1/4$. If the auctioneer repeats the same exact procedure (sending $|00\rangle$ probe-states and making measurements on the qubits thereafter) for $N$ iterations of the protocol without being caught, at the end of the $N^{\rm th}$ iteration, he would have learnt the bid values of both the bidders with probability $(1-(1/2)^N)^2$. This `learning-curve' of the corrupt auctioneer is the cyan line in Figure~\ref{fig:convergence}. Notice that even in this simple example, even with this simple `attack' on the protocol by the auctioneer, the auctioneer finds out the bid values of the bidders much faster than the convergence of the adiabatic auctions protocol. Hence, if the corrupt auctioneer is not caught, he can pretend to be running the auctions protocol as prescribed, and stop at the end of a stipulated number of iterations, and just decide the winner classically (and correctly). So, whereas on one hand nobody has a reason to suspect him, he learns all the bid-values of the losing bidders - hence compromising the privacy of losing bidders.

The corrupt auctioneer can actually do better, if he has access to the machinery to perform general $2$-qubit measurements. Lets say the auctioneer sends out probe states $|\psi_{\rm probe}\rangle = |00\rangle \otimes |00\rangle$ in each iteration of the protocol (as explained in the above paragraph), and receives a tensor product of the two bidders' bidding states (two out of the three states in Eq.~\ref{eq:biddingstates}). The auctioneer optimizes over all $3$-element $2$-qubit POVM elements $\Pi \equiv \left\{\Pi_1, \Pi_2, \Pi_3\right\}$ to distinguish between the three possible bidding-states (Eq.~\ref{eq:biddingstates}), and chooses the POVM measurement $\Pi^*$ that minimizes the probability of making a wrong decision $P_{\rm e}$ \footnote{For readers not familiar with quantum POVM measurements, a good reference is Helstrom's book on ``Quantum detection and estimation theory" \cite{helstrombook}. When POVM elements $\left\{\Pi_j\right\}_{j=1}^M$ are used to distinguish between pure states $\left\{|\psi_i\rangle\right\}_{i=1}^N$, the probability that the measurement outcome if $j$ given $|\psi_i\rangle$ was sent $p(j|i)$ is given by, $p(j|i) = \langle\psi_i|\Pi_j|\psi_i\rangle$. For distinguishing between mixed states $\left\{\rho_i\right\}_{i=1}^N$ using the same POVM elements, $p(j|i) = {\rm{Tr}}(\Pi_j\rho_i)$.  In order to numerically obtain the set of $3$-element POVM operators that minimize the probability of error in distinguishing between three pure states, we start with an arbitrary set of projective measurement operators (rank-1 operators corresponding to the self outer-products of a set of orthonormal vectors), and rotate the entire set of three states we want to distinguish between, around each one of the measurement vectors, minimizing probability of error in each iteration until convergence is reached. This greedy algorithm is not guaranteed to work for any set of pure states. But, \cite{helstrombook} gives a necessary criterion that a probability of error minimizing POVM must meet, which is a good check.}. Let's say the auctioneer uses $\Pi^*$ to measure both the bidders' bidding-states. The probability he learns the bid-values of both the bidders in one iteration is $(1-P_{\rm e})^2$. If the auctioneer repeats sending out $|\psi_{\rm probe}\rangle$ in each iteration of the protocol and keeps measuring the states he gets back from the bidders, using the optimum POVM operators $\Pi^*$, after $N$ such iterations, the probability that he learns both the bidders' bid-values correctly is given by $(1-P_{\rm e}^N)^2$. This `learning-curve' is the yellow line in Figure~\ref{fig:convergence} (For this example, the optimum measurement yields $P_{\rm e} = 0.1112$). Notice that using an optimum measurement, in this simple example, the auctioneer just needs about $4$ iterations to learn all the bid values almost perfectly. So, a corrupt auctioneer would just need to `spread-out' these cheating instants randomly in the $20$ iterations of the adiabatic protocol, and he can choose the winner by a classical search.

\item {\bf{Countering the attack: `locking operators'}} --- The essence of using `locking operators' is to rotate the bidding-states heavily towards the $|00\rangle$ state and away from the `price-state', so that it becomes harder for the corrupt auctioneer to distinguish between the three different bidding-states by using full-blown quantum measurements. To be more specific, consider an example in which one bidder bids $\$2$ and the other bidder bids $\$3$. Hence the two bidders' bidding operators $U_1$ and $U_2$ are such that $U_1|00\rangle = (|00\rangle + |10\rangle)/\sqrt{2}$ and $U_2|00\rangle = (|00\rangle + |11\rangle)/\sqrt{2}$. Now imagine that the two bidders come up with two secretly chosen unitary operators $V_1$ and $V_2$, such that

\begin{eqnarray}
V_1^\dagger\left(\frac{|00\rangle + |10\rangle}{\sqrt{2}}\right) = \alpha_1|00\rangle + \beta_1|10\rangle \nonumber \\
V_2^\dagger\left(\frac{|00\rangle + |11\rangle}{\sqrt{2}}\right) = \alpha_2|00\rangle + \beta_2|11\rangle \nonumber
\label{eq:lockingoperators}
\end{eqnarray}
with $|\alpha_i| \to 1$, for $i=1,2$. If a bidder always `locks' his bidding state using a $V$-operator with $\alpha \to 1$, the three possible bidding-states come very close to each other in the Hamming space, making detection by measurement harder. Now, consider expressing the final Hamiltonian $H_p$ of the adiabatic search in a basis defined by the unitary operator $V \equiv V_1 \otimes V_2$. So, the interpolated Hamiltonian is given by $H(f) = (1-f)H_b + fH_f = (1-f)UWU^\dagger + fVH_pV^\dagger$. Hence, the iterations of the adiabatic search will now look like:

\begin{eqnarray}
\ket{\Psi_s} &=& e^{-i\Delta{H(f)}}\ket{\Psi_{s-1}} \nonumber \\
&=& e^{-i\Delta{\left((1-f)H_b+fH_f\right)}}\ket{\Psi_{s-1}} \nonumber \\
&=& e^{-i\Delta{\left((1-f)UWU^{\dagger}+fVH_pV^\dagger\right)}}\ket{\Psi_{s-1}} \nonumber \\
&\approx& e^{-i\Delta{(1-f)}UWU^\dagger}e^{-i\Delta{fVH_pV^\dagger}}\ket{\Psi_{s-1}} \nonumber \\
&=& U\left(e^{-i\Delta{(1-f)W}}\right)U^{\dagger}Ve^{-i{\Delta}fH_p}V^\dagger\ket{\Psi_{s-1}} \nonumber \\
&=& UD(\Delta, 1-f)U^{\dagger}VP(\Delta, f)V^\dagger\ket{\Psi_{s-1}}
\label {eq:iterations_locking}
\end{eqnarray}

Observe that since the `locking operators' do not take the bidding-states out of the $2$-dimensional subspace in which the bidders placed their bids originally, by a careful choice of the locking operators (see Appendix A), the adiabatic search still remains in the initial search subspace. The only potential problem would occur, if the $g_{\rm min}$ of the new interpolated Hamiltonian $H(f) = (1-f)H_b + fH_f = (1-f)UWU^\dagger + fVH_pV^\dagger$ is very small. The convergence rate with this modified version of the protocol will depend upon the $g_{\rm min}$ for the new $H(f)$. The new `learning-curve' of the corrupt auctioneer (making symbol-by-symbol measurements in the computational basis) will now be given by $(1-|\alpha_1|^N)(1-|\alpha_2|^N)$, which will rise much slower with $N$ as compared to the `unprotected' case (which corresponds to $V = I_2 \otimes I_2)$, where $I_2$ is the $2 \times 2$ identity matrix). If the auctioneer knows before the fact, that such a protection mechanism might exist with bidders using `locking' operators $V_i$ without the auctioneer's prior knowledge, the corrupt auctioneer would be deterred to use the above cheating mechanism as he might be afraid of not being able to learn the bid values of the bidders by the end of the $S$ iterations, with significant probability - hence making the incorrect (non-revenue-maximizing) allocation. Observe that using this simple technique, the `learning-curve' of the corrupt auctioneer who only has access to single qubit projective measurements, can be slowed down by a huge amount. But, if the auctioneer chooses to use a more complicated two-qubit POVM, then the advantage of using this strategy to counter the attack is less pronounced. For a concrete example on how the auctioneer's `learning-curve' is slowed down by using `locking-operators', see Figure~\ref{fig:locking}.

\begin{figure}
\begin{center}
\includegraphics[height=7cm,width=9cm,angle=0]{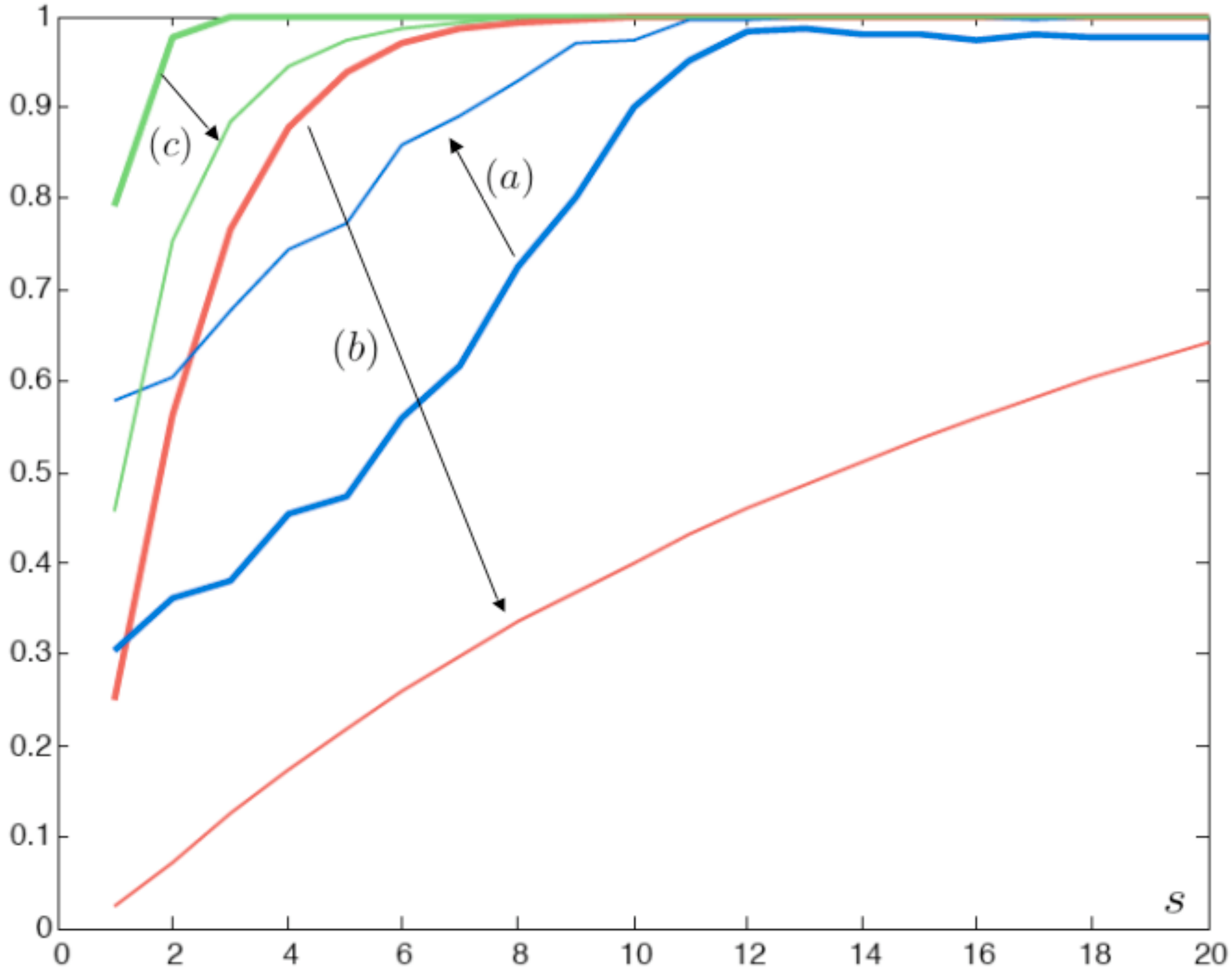}
\end{center}
\caption{This figure depicts how quickly a corrupt auctioneer can learn the bid values of all the bidders by repeatedly secretly sending out fresh copies of initialized qubits to all bidders and making projective measurements on the qubits he gets back from the bidders thereafter. The thick lines correspond to (a) Blue line: convergence of the adiabatic protocol, (b) Red line: learning curve of the corrupt auctioneer using single-qubit measurements on all qubits in the computational basis, and (c) Green line: the learning curve of the corrupt auctioneer who uses the most optimum two-qubit POVM to distinguish between possible bidding states. The thin curves are the corresponding plots when both bidders use locking operators. Bidder 1 bids $\$3$, and $\alpha_1 = 0.9$, and bidder 2 bids $\$2$ and $\alpha_2 = 0.7$. We see that in this example, not only are the learning curves of the corrupt auctioneer `slowed down' by using secret locking operators by the bidders, the convergence of the adiabatic protocol has become faster.}
\label{fig:locking}
\end{figure}
\end{enumerate}

\subsection{Attack using spurious Hamiltonian}

\begin{enumerate}

\item {\bf {The attack mechanism: `revealing-states and spurious $H_p$'}} --- The choice of the final Hamiltonian $H_p$ depends on the payoff structure of the auctions, and thus must be left to the discretion of the auctioneer. The revenue-maximizing $H_p$ we described in \ref{example} assigns zero-payoff to what we termed as `infeasible' states. Imagine, that the auctioneer uses a final Hamiltonian $H_p = {\rm diag}\left\{0, -1, -2, -3, -1, -2, -3, -4, -2, -3, -4, -5,\right.$ $\left. -3, -4, -5, -6\right\}$. A `revealing' state is a state that contains information about bid values of both bidders, for example $|1011\rangle$ is a `revealing' state. The following payoff table assigns maximum payoff values (thus minimum eigenvalues) to the revealing states, so that the adiabatic search always converges to the revealing state in the search-subspace. The payoff values for this spurious Hamiltonian are the sum of the dollar-values of the two bidders' basis states. The corrupt auctioneer would thus always be able to converge to the revealing-state in the superposition of four states, as it will be the `ground' state of the final Hamiltonian in the search subspace. And when the auctioneer does the final measurement on the qubits after $S$ iterations of the adiabatic algorithm, he learns both the bid values and then he can decide the winner classically. This form of attack is seemingly very hard to detect, and as the auctioneer is still running the adiabatic search, there is no reason for suspicion. See Table~\ref{table:payoff_spurious} for the payoff values used by the corrupt auctioneer. As an example, if the bidders bid $\$2$ and $\$3$ respectively, the search would converge to the state $|1011\rangle$ rather than $|0011\rangle$, so the auctioneer would first learn both the bid values by making a measurement on all $4$ qubits, and then he'ld pick the winner classically. Figure~\ref{fig:convergence_spuriousHp} shows the convergence rate of the corrupt auctioneer to the `revealing-state' $|1011\rangle$ when one bidder bids $\$2$ and the other $\$3$. Figure~\ref{fig:eigenvalues_spuriousHp} shows the eigenvalues of the interpolated Hamiltonian with the spurious $H_p$.

\begin{table}
\tbl{A corrupt auctioneer trying to pry on the bid-values of even the losing bidders, might choose a different payoff metric, which would reveal the bid-values of all the bidders. Consider the following auction payoffs for two players $B_1$ and $B_2$ bidding on one item. Each player has two qubits. The `allocation' $\ket{x} = \ket{q_1q_2}_{B_1} \ket{q_3 q_4}_{B_2}$ has a payoff given by $F(x) =  -\bra{x}H_p\ket{x}$, where $H_p$ is a diagonal matrix that constitutes the final Hamiltonian of the adiabatic search. A `revealing' state is state that contains information about bid values of both bidders, for example $|1011\rangle$ is a `revealing' state. The following payoff table assigns maximum payoff values to the revealing states, so that the adiabatic search always converges to the revealing state in the search-subspace. The payoff values for this spurious Hamiltonian are the sum of the dollar-values of the states $\ket{x} =\ket{q_1q_2}_{B_1}$ and $ \ket{q_3 q_4}_{B_2}$.}
{\begin{tabular}{|ccc|}
\hline
$\ket{q_1 q_2}_{B_1}$ & $\ket{q_3 q_4}_{B_2}$ & $F(x)$ \\
\hline
 00 &  00 &  0  \\
 00 &  01 &  1  \\
 00 &  10 &  2  \\
 00 &  11 &  3  \\
 01 &  00 &  1  \\
 01 &  01 &  2  \\
 01 &  10 &  3  \\
 01 &  11 &  4  \\
 10 &  00 &  2  \\
 10 &  01 &  3  \\
 10 &  10 &  4  \\
 10 &  11 &  5  \\
 11 &  00 &  3  \\
 11 &  01 &  4  \\
 11 &  10 &  5  \\
 11 &  11 &  6  \\
\hline
\end{tabular}}
\label{table:payoff_spurious}
\end{table}

\begin{figure}
\begin{center}
\includegraphics[height=7cm,width=9cm,angle=0]{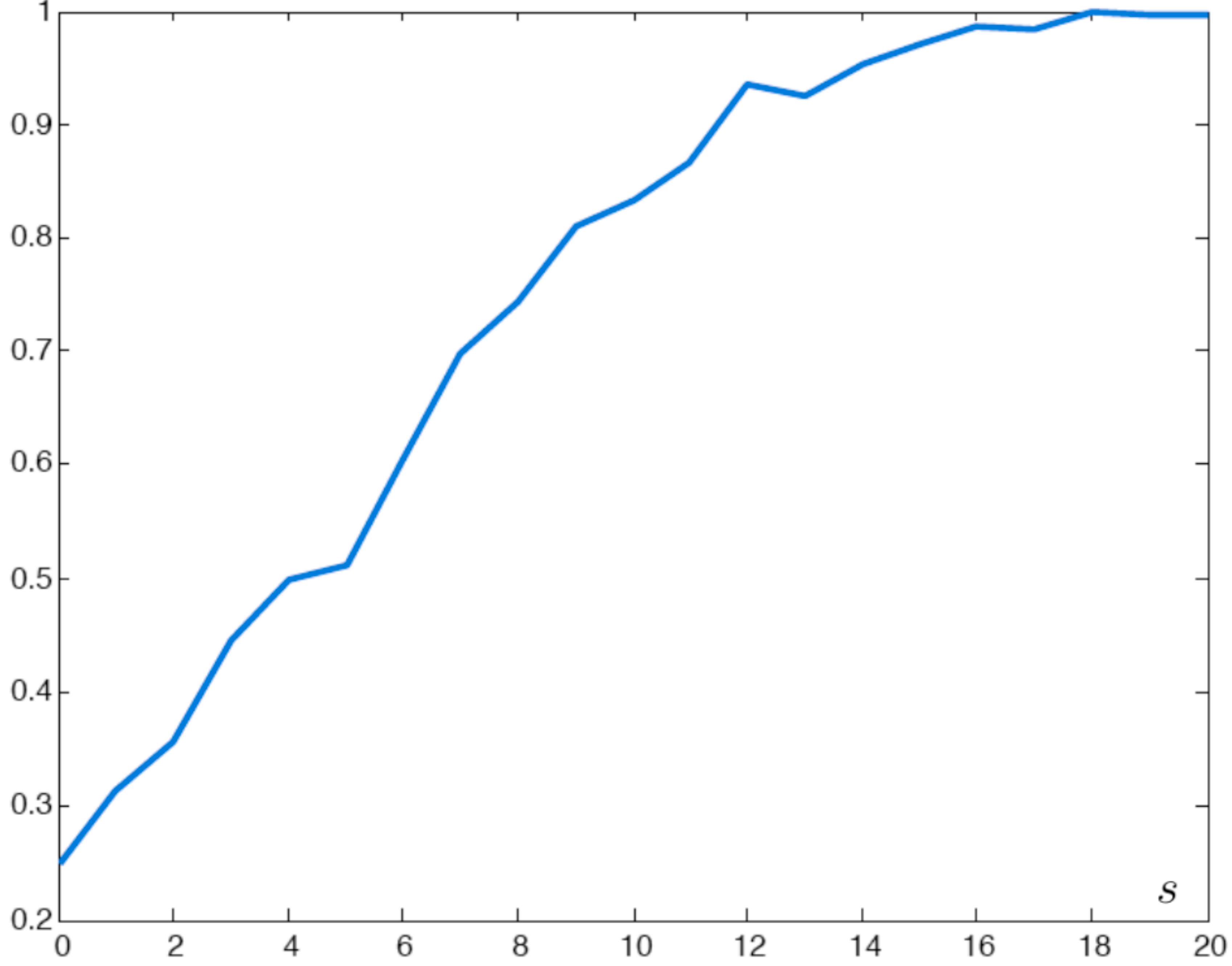}
\end{center}
\caption{This figure shows the convergence rate of the adiabatic search (using the spurious $H_p$ of the corrupt auctioneer) to the `revealing-state' $|1011\rangle$ when one bidder bids $\$2$ and the other $\$3$. After the adiabatic search is over, the auctioneer makes a measurement and finds out both the bid values. He decides the winner by classical search.}
\label{fig:convergence_spuriousHp}
\end{figure}

\begin{figure}
\begin{center}
\includegraphics[height=7cm,width=9cm,angle=0]{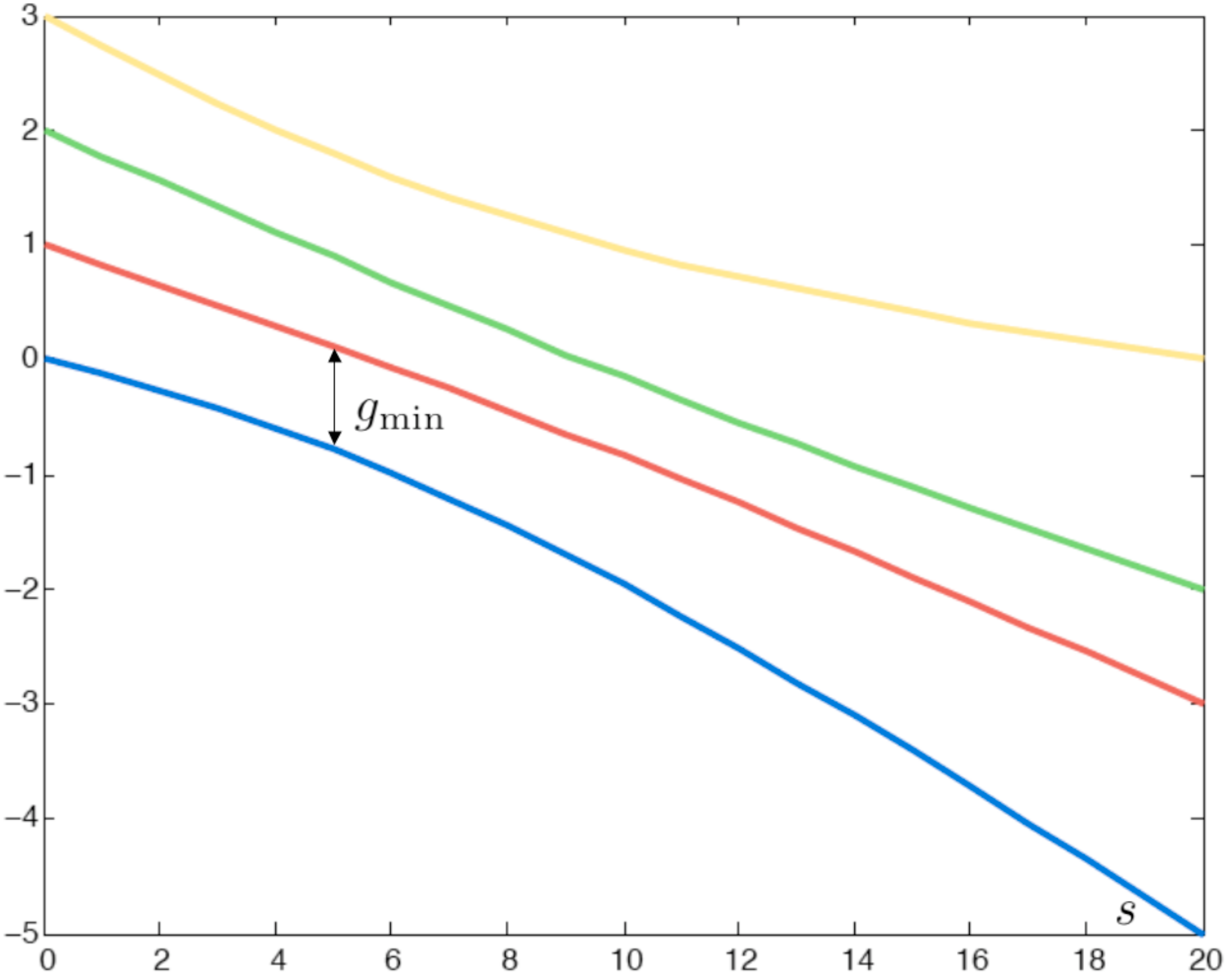}
\end{center}
\caption{A plot of the eigenvalues of the interpolated Hamiltonian $H(f)$ as a function of the iteration number $s$, where $f=s/S$ and $S=20$ is the total number of iterations. These eigenvalues correspond to the case when the spurious final Hamiltonian $H_p$ is used by a corrupt auctioneer, and when the bidders bid $\$2$ and $\$3$.}
\label{fig:eigenvalues_spuriousHp}
\end{figure}

\item {\bf{Countering the attack: `removal of revealing states by collusion'}} --- The above mode of attack can be countered if the bidders can somehow get rid of the `revealing state' from their joint bidding states. This can be achieved if the suspecting bidders collude with each other in order to prepare their bidding-states jointly. Lets say that bidder-1 bids $\$2$ on the item and bidder-2 bids $\$3$. Their bidding operators are $U_1$ and $U_2$ respectively. All they need to do is to agree to come together with their bidding operators $U_1$ and $U_2$ respectively, and put them together with a rotation operator ${\cal R}(\sqrt{2/3},\sqrt{1/3})$ and the controlled-$0$ operations as shown in Figure~\ref{fig:collusion}. The rotation operator rotates the uniform superposition state $(|00\rangle+|10\rangle)/\sqrt{2}$ to the state $\sqrt{2/3}|00\rangle + \sqrt{1/3}|10\rangle$. The controlled-$0$ operation means that $U_2$ is applied to the third and the fourth qubits (the qubits that belong to bidder-2), only when both the qubits of the first bidder are in their $|0\rangle$ states. So, now the joint unitary operator $U$ of the bidders is no longer a tensor product of $U_1$ and $U_2$. In each iteration of the adiabatic algorithm, the bidders would need to use the additional apparatus in Figure~\ref{fig:collusion} to apply the $U$ operator jointly. Note that now, even with the spurious final Hamiltonian $H_p$, the auctioneer will not be able to read both the bid values. But the auctioneer will still make the right decision, because the state $|0011\rangle$ has the highest payoff (least eigenvalue) out of the three states in the superposition in Figure~\ref{fig:collusion} (ii).

\begin{figure}
\begin{center}
\includegraphics[height=8cm,width=12cm,angle=0]{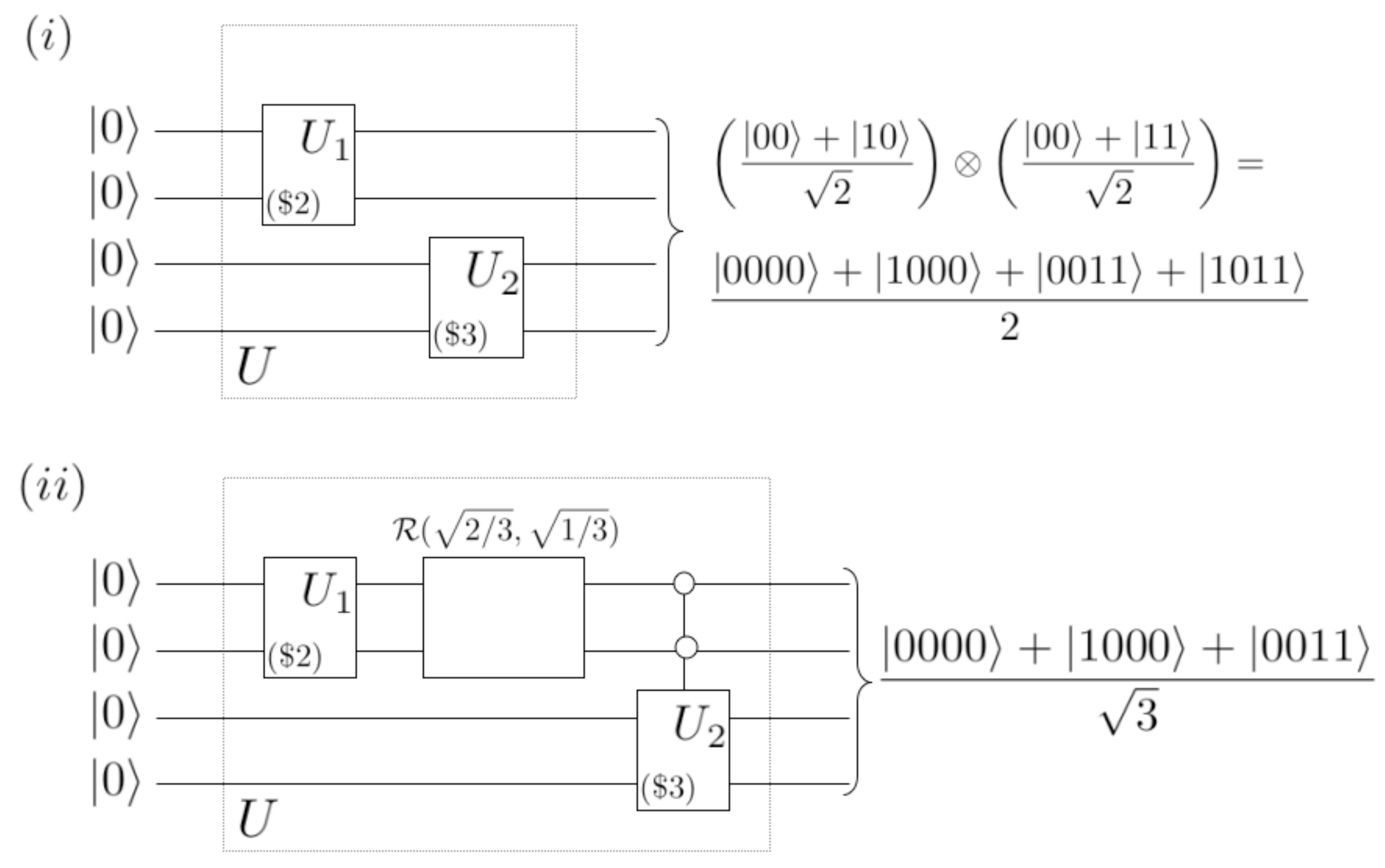}
\end{center}
\caption{This figure shows how the bidders can collude to get rid of the `revealing-state' (in this case $|1011\rangle$) from the joint bidding state. Note that in order to do that, the bidders do not need to reveal their bid values to each other. They just need to agree to come together with their bidding operators $U_1$ and $U_2$ respectively, and put them together with a rotation operator ${\cal R}(\sqrt{2/3},\sqrt{1/3})$ and the controlled-$0$ operations as shown in the figure. The rotation operator rotates the uniform superposition state $(|00\rangle+|10\rangle)/\sqrt{2}$ to the state $\sqrt{2/3}|00\rangle + \sqrt{1/3}|10\rangle$. The controlled-$0$ operation means that $U_2$ is applied to the third and the fourth qubits (the qubits that belong to bidder-2), only when both the qubits of the first bidder are in their $|0\rangle$ states.}
\label{fig:collusion}
\end{figure}

\end{enumerate}

\section{Conclusions}

Quantum information technology offers a new paradigm for various
economic mechanisms, such as the auction protocol discussed in
this paper. It can improve on pre-existing protocols by making them
more secure, more efficient, or by adding some novel privacy aspect to them.
More generally, it provides a new framework for economic games,
with new optimum strategies and new ways of cheating for the
participants --- with possible detrimental effects on economic
outcomes. In the context of auctions, the classical economic
analysis usually is not concerned about privacy of the bidders,
and focuses exclusively on the bidders' behavior under the
assumption that the auctioneer performs the auction as specified
in the protocol. However, playing the auction game with quantum
hardware can improve the protocol by guaranteeing the privacy of
the losing bidders. This added feature implicitly treats the
information about the bids as a valuable resource, and introduces
a new incentive for the auctioneer to play the game dishonestly in
order to learn that information. Therefore, the economic analysis
of the auction game has to be carefully reexamined.

Our discussion focused on methods that the auctioneer could use to
learn more information about the various bids than the honest
protocol allows. In these scenarios, we showed that the time
required for the auctioneer to learn extra information about the
values of the bids could be made significantly shorter than the time
required to complete the honest auction protocol --- especially when
the auctioneer has access to a quantum computer --- so that the
bidders would have no way to detect the dishonest behavior. In
this case, the auction would be useful only to the extent the
bidders trust the auctioneer to behave correctly, although it would still
offer advantages over classical auctions, such as a unique way for
the bidders to correlate their bids. To reduce this required
trust, we have described techniques by which the bidders could
significantly slow down the rate at which illicit information is learned by
the auctioneer during the protocol. In practice, whether
these techniques are sufficient to ensure the auctioneer's honest
behavior depends on the context in which the auction is used. For
instance, the value that the auctioneer places on a reputation
for correct behavior (e.g., to attract future business) could be compared
to the potential value of learning more about bidders'
preferences, or the extent to which the auctioneer is risk averse
(in the context of possible criminal prosecution) could be considered. This
observation also indicates a choice in the search algorithm --- in
some cases it may be better to run the search for a fewer number of steps, or
use alternate heuristic search methods, which give the auctioneer
fewer opportunities to probe bidder states, even at the cost of
somewhat increased probability the highest bidder does not win due
to error in the search.

\section*{Acknowledgments}
The authors acknowledge helpful discussions with Kay-Yut Chen (HP Labs), Phil Kuekes (HP Labs) and Pavithra Harsha (MIT). This work was supported by DARPA and the Army Research Office through ARO contract \#W911NF0530002. This article does not necessarily reflect the position or the policy of the Government funding agencies, and no official endorsement of the views contained herein by the funding agencies should be inferred.

\appendix
\section{Comments on subspace search}

As we observe in footnote~\ref{footUj}, the bidding operators $U_j$ for each bidder can be constructed in many different ways, leading to different bidding operators $U$ that can implement the adiabatic search for the auction protocol described in the paper. For any arbitrary selection of the bidding operator $U$, it is NOT necessarily true that the search procedure (as described in Eq.~\eqref{eq:iterations}) does not ever go out of the initial subspace spanned by the vectors describing the bidding states of all the bidders, $|\Psi_0\rangle$. In this appendix, we show that the search indeed does remain in the span of the initial bidding states, if the bidders create the bidding operators by a special Hadamard-like construction.

For sake of simplicity, we would like to allude to the $4$-qubit two-bidder ``toy example" that we refer to, in Section 3 of the paper. Let us consider the construction of the bidding operators $U^{(1)}, U^{(2)}$, and $U^{(3)}$ shown in Fig.~\ref{fig:unitops}. that create the three bidding states $|\psi_i\rangle = U^{(i)}|00\rangle$, by action on the initially distributed state $|00\rangle$:

\begin{eqnarray}
|\psi_1\rangle = \frac{|00\rangle + |01\rangle}{\sqrt{2}}  \nonumber \\
|\psi_2\rangle = \frac{|00\rangle + |10\rangle}{\sqrt{2}}  \nonumber \\
|\psi_3\rangle = \frac{|00\rangle + |11\rangle}{\sqrt{2}}.
\label{eq:biddingstates}
\end{eqnarray}

The conditions specified by Eqns.~\eqref{eq:biddingstates} only specify the first column of each unitary operator $U^{(i)}$ in the standard Kronecker basis. Hence, there can be multiple possible ways to extend the first column into a complete-orthonormal (CON) set of (four) basis vectors (4-length column vectors) to construct the unitary matrices. We use the standard Gram-Schmidt orthonormalization process to extend the first column to a Hadamard-like construction. These constructions can be summarized in quantum-circuit notation by the circuits shown in Fig.~\ref{fig:unitops}. The Hadamard matrix is given by 

\begin{equation}
H = \frac{1}{\sqrt{2}}\left[
\begin{array}{cc}
1 & 1\\
1 & -1
\end{array}
\right].
\end{equation}
Using this construction (as can be also verified by explicit calculations of the output states for the circuits in Fig.~\ref{fig:unitops}), we obtain the following unitary bidding operators $U^{(i)}$:

\begin{eqnarray}
U^{(1)} &=& \frac{1}{\sqrt{2}}\left[
\begin{array}{cccc}
1 & 1 & 0 & 0\\
1 & -1 & 0 & 0\\
0 & 0 & 1 & 1\\
0 & 0 & 1 & -1
\end{array}
\right]\\
U^{(2)} &=& \frac{1}{\sqrt{2}}\left[
\begin{array}{cccc}
1 & 0 & 1 & 0\\
0 & 1 & 0 & 1\\
1 & 0 & -1 & 0\\
0 & 1 & 0 & -1
\end{array}
\right]\\
U^{(3)} &=& \frac{1}{\sqrt{2}}\left[
\begin{array}{cccc}
1 & 0 & 1 & 0\\
0 & 1 & 0 & 1\\
0 & 1 & 0 & -1\\
1 & 0 & -1 & 0
\end{array}
\right]
\end{eqnarray}

Note that such a Hadamard-like construction of unitary bidding operators is always possible even in a more general case of more than two bidders and more number of qubits. The overall bidding operator is given by $U = U_1 \otimes U_2$, where $U_1$ and $U_2$ are bidding operators of the two bidders. The adiabatic iterations are given by:
\begin{equation}
|{\Psi_s}\rangle = UD(\Delta, 1-f)U^{\dagger}P(\Delta, f)|{\Psi_{s-1}}\rangle,
\label{eq:search2}
\end{equation}
where $D(\Delta, 1-f) = e^{-i\Delta(1-f)W}$ and $P(\Delta, f) = e^{-i\Delta{f}H_p}$ are diagonal unitary matrices with unit-magnitude phase-terms as their diagonal entries. We will see in the following subsection, that given the above construction of the $U$'s, the adiabatic iterations preserve the initial search subspace. As an example, if bidder 1 bids $\$2$, and bidder 2 bids $\$3$, i.e. $U_1 = U^{(2)}$, $U_2 = U^{(3)}$, and $U = U^\dagger = U_1 \otimes U_2$, then at each iteration step $s$, the state $|\psi_s\rangle$ is in the span of the vectors $|0000\rangle$, $|0011\rangle$, $|1000\rangle$, and $|1011\rangle$. 

\subsection*{Hadamard construction of $U$ preserves search subspace}

Our goal in this section is to show that the Hadamard-like construction of the bidders' unitary operator $U$ described above preserves the search subspace. In other words, we want to show that, with our construction of $U$, $UD(\Delta, 1-f)U^\dagger$ preserves the subspace spanned by the initial basis-states in the uniform superposition $|\Psi_0\rangle$.
 
In order to start the search in the right state, we want $U$ to map the initial state $|\Psi_{\rm init}\rangle = |0000\rangle$ to the uniform superposition, $|\Psi_0\rangle = (1/2)(|{\bf 0}{\bf 0}\rangle+|{\bf 0}{\bf b_2}\rangle+|{\bf b_1}{\bf 0}\rangle+|{\bf b_1}{\bf b_2}\rangle$\footnote{Here we use the shorthand notation $|{\bf 0}\rangle$ to denote the two-qubit state $|00\rangle$. $|{\bf b_1}\rangle$ and $|\bf b_2\rangle$ denote one out of $3$ possible non-zero bid values ($|01\rangle$, $|10\rangle$ or $|11\rangle$) for bidder 1 and bidder 2 respectively.}. An equivalent way of saying this is that we want ${U^\dagger}$ to map the state $(1/2)(|{\bf 0}{\bf 0}\rangle+|{\bf 0}{\bf b_2}\rangle+|{\bf b_1}{\bf 0}\rangle+|{\bf b_1}{\bf b_2}\rangle$ onto $|{\bf 0}{\bf 0}\rangle$. Then $UD(\Delta, 1-f)U^\dagger$ sends $(1/2)(|{\bf 0}{\bf 0}\rangle+|{\bf 0}{\bf b_2}\rangle+|{\bf b_1}{\bf 0}\rangle+|{\bf b_1}{\bf b_2}\rangle$ onto itself. Because $P(\Delta, f) = e^{-i\Delta{f}H_p}$ is a diagonal matrix, it preserves ${span}(|{\bf 0}{\bf 0}\rangle, |{\bf 0}{\bf b_2}\rangle, |{\bf b_1}{\bf 0}\rangle, |{\bf b_1}{\bf b_2}\rangle)$. So, all we want now is that $UD(\Delta, 1-f)U^\dagger$ also preserves the above initial bidding subspace. 

As $UD(\Delta, 1-f)U^\dagger$ sends $(1/2)(|{\bf 0}{\bf 0}\rangle+|{\bf 0}{\bf b_2}\rangle+|{\bf b_1}{\bf 0}\rangle+|{\bf b_1}{\bf b_2}\rangle$ onto itself, one way to construct a $U$ that also preserves the subspace given by ${span}(|{\bf 0}{\bf 0}\rangle, |{\bf 0}{\bf b_2}\rangle, |{\bf b_1}{\bf 0}\rangle, |{\bf b_1}{\bf b_2}\rangle)$, is to enforce $U^\dagger$ to map $3$ mutually orthogonal linear combinations of the vectors $\left\{(|{\bf 0}{\bf 0}\rangle, |{\bf 0}{\bf b_2}\rangle, |{\bf b_1}{\bf 0}\rangle, |{\bf b_1}{\bf b_2}\rangle\right\}$ (each of which are also orthogonal to $(1/2)(|{\bf 0}{\bf 0}\rangle+|{\bf 0}{\bf b_2}\rangle+|{\bf b_1}{\bf 0}\rangle+|{\bf b_1}{\bf b_2}\rangle$) onto orthogonal eigenstates of operator $D(\Delta, 1-f)$ (which are just the computational basis states). This way, $UD(\Delta, 1-f)U^\dagger$ will map each one of these linear combinations onto itself, albeit with an added phase factor coming from the corresponding eigenvalue of $D(\Delta, 1-f)$. Moreover, now because $UD(\Delta, 1-f)U^\dagger$ maps an orthonormal basis of the subspace ${span}(|{\bf 0}{\bf 0}\rangle, |{\bf 0}{\bf b_2}\rangle, |{\bf b_1}{\bf 0}\rangle, |{\bf b_1}{\bf b_2}\rangle)$ onto the subspace, it would map any arbitrary linear combination of the vectors $\left\{|{\bf 0}{\bf 0}\rangle, |{\bf 0}{\bf b_2}\rangle, |{\bf b_1}{\bf 0}\rangle, |{\bf b_1}{\bf b_2}\rangle\right\}$ to another linear combination of these same vectors. Thus, with such a choice of $U$, the adiabatic search process described by Eq.~\eqref{eq:search2} will never leave the initial `bidding subspace'. 

The Hadamard-like construction we are implementing is one example of this method, where we choose the $3$ remaining linear combinations to be:

\begin{eqnarray}
(1/2)(|{\bf 0}{\bf 0}\rangle+|{\bf 0}{\bf b_2}\rangle-|{\bf b_1}{\bf 0}\rangle-|{\bf b_1}{\bf b_2}\rangle,&& \nonumber \\
(1/2)(|{\bf 0}{\bf 0}\rangle-|{\bf 0}{\bf b_2}\rangle+|{\bf b_1}{\bf 0}\rangle-|{\bf b_1}{\bf b_2}\rangle,&& \quad \text{and} \nonumber \\
(1/2)(|{\bf 0}{\bf 0}\rangle-|{\bf 0}{\bf b_2}\rangle-|{\bf b_1}{\bf 0}\rangle+|{\bf b_1}{\bf b_2}\rangle.&& \nonumber
\end{eqnarray}
We could also pick some other $3$ linear combinations, but the above work as desired by the protocol. 

\subsection*{Choice of locking operators}

A similar consideration as above arises for the modified adiabatic search protocol when the bidders use `locking operators' \eqref{eq:iterations_locking}. By a specific choice of the locking operators, we can again always ensure that the adiabatic iterations remain in the initial search subspace. Referring to the specific case elaborated in the paper (Eq.\eqref{eq:lockingoperators}), we construct the matrices $V_1$, and $V_2$ in the following manner (in the standard computational Kronecker basis):

\begin{eqnarray}
V_1 = V_1^\dagger &=& \left[
\begin{array}{cccc}
\cos{\theta_1} & 0 & \sin{\theta_1} & 0\\
0 & \cos{\theta_1} & 0 & \sin{\theta_1}\\
\sin{\theta_1} & 0 & -\cos{\theta_1} & 0\\
0 & \sin{\theta_1} & 0 & -\cos{\theta_1}
\end{array}
\right] \quad \text{and}\\
V_2 = V_2^\dagger &=& \left[
\begin{array}{cccc}
\cos{\theta_2} & 0 & 0 & \sin{\theta_2}\\
0 & \cos{\theta_2} & \sin{\theta_2} & 0\\
0 & \sin{\theta_2} & -\cos{\theta_2} & 0\\
\sin{\theta_2} & 0 & 0 & -\cos{\theta_2}
\end{array}
\right],
\end{eqnarray}
such that,
\begin{eqnarray}
\cos{\theta_i} + \sin{\theta_i} &=& \alpha_i, \quad \text{and}\\
\cos{\theta_i} - \sin{\theta_i} &=& \beta_i,
\end{eqnarray}
for $i =1, 2$. With the above choice of the `locking operators', it is again easy to see why the iterative search procedure (Eq.~\eqref{eq:iterations_locking}) will never go outside the subspace spanned by the allocation vectors in the initial state.

\end{document}